\documentclass[aps,pra,superscriptaddress,showpacs,twocolumn]{revtex4-1}
\usepackage{bm}
\usepackage{amsmath}
\usepackage{float}
\allowdisplaybreaks
\usepackage{longtable}
\usepackage{nicefrac}
\usepackage{dcolumn}
\newcolumntype{.}{D{x}{}{-1}}

\newcommand*{\centt}[1]{\multicolumn{1}{c}{#1}}
\newcolumntype{w}[1]{D{.}{.}{#1}}

\newcommand{\bfr}{\vec{r}}

\newcommand{\bl}{\ln k_0}
\usepackage{graphicx}
\usepackage{xcolor}
\def\mynote1#1{{\color{blue}{\textsc{\footnotesize Question/Comment:} #1}}}
\def\mynote2#1{{\color{magenta}{#1}}}
\newcommand{\lbr}{\langle}
\newcommand{\rbr}{\rangle}

\begin{document}
\title{Atomic structure calculations of helium with correlated exponential functions}

\author{Vladimir A. Yerokhin}
\affiliation{Center for Advanced Studies, Peter the Great St.~Petersburg Polytechnic University,
Polytekhnicheskaya 29, 195251 St.~Petersburg, Russia}

\author{Vojt\v{e}ch Patk\'o\v{s}}
\affiliation{Faculty of Mathematics and Physics, Charles University,  Ke Karlovu 3, 121 16 Prague
2, Czech Republic}

\author{Krzysztof Pachucki}
\affiliation{Faculty of Physics, University of Warsaw,
             Pasteura 5, 02-093 Warsaw, Poland}

\date{\today}

\begin{abstract}

The technique of quantum electrodynamics (QED) calculations of energy levels in the helium atom
is reviewed. The calculations start with the solution of the Schr\"odinger equation and account
for relativistic and QED effects by perturbation expansion in the fine-structure constant
$\alpha$. The nonrelativistic wave function is represented as a linear combination of basis
functions depending on all three interparticle radial distances, $r_1$, $r_2$ and $r =
|\vec{r}_1-\vec{r}_2|$. The choice of the exponential basis functions of the form $\exp(-\alpha
r_1 -\beta r_2 -\gamma r)$ allows us to construct an accurate and compact representation of the
nonrelativistic wave function and to efficiently compute matrix elements of numerous singular
operators representing relativistic and QED effects. Calculations of the leading QED effects of
order $\alpha^5m$ (where $m$ is the electron mass) are complemented with the systematic
treatment of higher-order $\alpha^6m$ and $\alpha^7m$ QED effects.
\end{abstract}

\maketitle

\section{Introduction}

The helium atom is the simplest many-body atomic system in the nature. Since the advent of
quantum mechanics, helium was used as a benchmark case for developing and testing various
calculational approaches of many-body atomic theory. Today, the nonrelativistic energy of
various helium electronic states can be computed with an essentially arbitrary numerical accuracy
\cite{schwartz:06,aznabaev:18}. The same holds also for the leading-order relativistic
correction. Subsequently, the quantum electrodynamics (QED) effects in the atomic structure of helium
can be clearly identified and studied by comparison of theoretical predictions with the large body
of available experimental data. Experimental investigations of helium spectra have progressed
rapidly over the years, recently reaching the precision of a few tens of Hertz
\cite{kato:18}.

For light atomic systems such as helium, relativistic and QED corrections to energy levels can be
systematically accounted for by the perturbation expansion in the fine-structure constant
$\alpha$. The starting point of the expansion is the nonrelativistic energy of order
$\alpha^2\,m$ ($=$ 2\,Ry, where $m$ is the electron mass and Ry is the Rydberg energy). The
leading relativistic correction is of order $\alpha^4\,m$, whereas QED effects enter first in
order $\alpha^5\,m$. A large body of work has been done in recent years to calculate QED effects
in helium spectra. Extensive calculations of helium energies were accomplished by Gordon Drake
{\em et al.} \cite{drake:98:cjp,morton:06:cjp,drake:01}. Their calculations are complete through
order $\alpha^5\,m$ and approximately include some higher-order QED effects. The next-order
$\alpha^6\,m$ QED correction was for a long time known only for the fine-structure intervals
\cite{lewis:78,zhang:96:prl}. For individual energy levels, these effects were derived and
calculated numerically by one of us (K.P.)
\cite{pachucki:02:jpb,pachucki:06:hesinglet,pachucki:06:he}. The higher-order $\alpha^7\,m$ QED
effects were evaluated by us first for the fine structure
\cite{pachucki:06:prl:he,pachucki:09:hefs,pachucki:10:hefs} and just recently for the triplet $n
= 2$ states of helium \cite{yerokhin:18:betherel,patkos:20,patkos:21}.

The purpose of this article is to review and systematize the technique of calculations of the
helium atomic structure, developed in numerous investigations over the last three decades. The
starting point of the calculations is the Schr\"odinger equation, which is solved variationally
after expanding the wave function into a finite set of explicitly-correlated basis functions
depending on all three interparticle radial distances. It has been known for a long time
\cite{hylleraas:29} that inclusion of the interelectronic distance explicitly into the basis set
is crucially important for constructing an accurate representation of the two-electron wave
function. Moreover, it has also long been recognized \cite{kato:57:cusp} that an accurate
wave-function representation should satisfy the so-called cusp conditions at the two-particle
coalescence points $|\vec{r}_i-\vec{r}_j| = 0$. The cusp condition is expressed \cite{pack:66},
after averaging over angles and for the singlet states, as
\begin{align}
\psi^{(S)}(r) \underset{r \to 0}{=} \psi^{(S)}(0)\, \big( 1 + \lambda\,r\big) + O(r^2)\,,
\end{align}
where $r$ is an interparticle distance and the parameter $\lambda = 1/2$ for the
electron-electron and $\lambda = -Z$ for the electron-nucleus cusp (where $Z$ is the nuclear
charge number).

The two most succesful basis sets used in the literature for high-precision calculations
of the atomic structure of helium are: the Hylleraas basis set adopted by Drake {\em et al.}
\cite{drake:98:cjp,morton:06:cjp,drake:01} and the exponential basis set put forward by Korobov
\cite{korobov:00,korobov:02} and used in numerous calculations of our group. Both these basis
sets are explicitly correlated and are able to reproduce the cusp conditions with great accuracy.
In the present work we will concentrate on the exponential basis set, because only this basis has been
successfully used in calculations of higher-order QED effects so far.
\section{Wave functions}

The spatial wave function $\psi_{ LM_L}$  with a specified total angular momentum $L$ and its
momentum projection $M_L$ for a two-electron atom is standardly represented as
\begin{align}\label{eq:2}
\psi_{ LM_L} = \sum_{l_1l_2} f_{l_1l_2}(r_1,r_2,r)\, Y_{LM_L}^{l_1l_2}(\hat{r}_1,\hat{r}_2)\,,
\end{align}
where $f_{l_1l_2}$ is the radial part of the wave function, $\vec r = \vec r_1 - \vec r_2$, and $\hat
r = \vec r/r$. Furthermore, $Y_{LM_L}^{l_1l_2}$ are the bipolar spherical harmonics,
\begin{align}
Y_{LM_L}^{l_1l_2}(\hat{r}_1,\hat{r}_2) = \sum_{m_1m_2} \lbr l_1m_1l_2m_2| LM_L\rbr\,
Y_{l_1m_1}(\hat{r}_1)\,Y_{l_2m_2}(\hat{r}_2)\,,
\end{align}
where $\lbr l_1m_1l_2m_2|lm\rbr$ is the Clebsch-Gordan coefficient and $Y_{lm}$ are the spherical
harmonics. We stress that the radial part of the wave function is assumed to be explicitly
correlated, i.e., the function $f$ depends on all interparticle distances, $r_1$, $r_2$, and $r$.
In this case, the sum over $l_1$ and $l_2$ in Eq.~(\ref{eq:2}) is restricted \cite{schwartz:61}
by two conditions
\begin{align}
{\rm (A):}\ l_1+l_2 = L\,,  \ \ {\rm or}\ \ {\rm (B):} \  l_1+l_2 = L + 1\,,
\end{align}
which lead to wave functions of different parities $(-1)^{l_1+l_2}$. The bipolar spherical
harmonics are usually handled in the spherical coordinates using the apparatus of Racah algebra,
see, e.g., Ref.~\cite{drake:78}. We find, however, that calculations with explicitly correlated
functions are more conveniently performed in Cartesian coordinates. One of the reasons is that
the action of numerous momentum operators encountered in calculations is most easily evaluated in
the Cartesian coordinate system. The corresponding calculations can easily be automatized and
performed with the help of systems of symbolic computations.

For this purpose, the expansion of the wave function is more conveniently made  in terms
of the bipolar solid harmonics. In order to define them, we start with the solid harmonics,
\begin{align}
{\cal Y}_{LM}(\vec r) = \sqrt{4\,\pi}\,A_L\,r^L\,Y_{LM}(\hat r)\,,
\end{align}
where the normalization coefficient $A_L$ is fixed below. The solid harmonics obey the following
summation rule,
\begin{align}
&\frac{1}{2\,L+1}\,\sum_{M=-L}^{L}\,{\cal Y}^*_{LM}(\vec{r\,}')\,{\cal Y}_{LM}(\vec r) = A_L^2\,r'^L\,r^L\,P_L(\hat r'\cdot\hat r)
\nonumber \\ &\ \hskip1cm = (r'^i\,r'^j\,r'^k\ldots)^{(L)}\,(r^i\,r^j\,r^k\ldots)^{(L)}\,, \label{sum1}
\end{align}
where $(r^i\,r^j\,r^k\ldots)^{(L)}$ is a traceless and symmetric tensor of the order $L$
constructed from components of the vector $\vec r$ with Cartesian indices $i,j,k\ldots$.
and the summation over these Cartesian indices is implicit. The last
equation determines $A_L$, which is related to the coefficient of $x^L$ in the Legendre polynomial
$P_L(x)$, specifically,
\begin{align}
 A_L^{-2} = \frac{1}{2^L}\,{2\,L \choose L}\,.
\end{align}
We now define the bipolar solid harmonics ${\cal Y}_{LM_L}^{l_1l_2}$ as
\begin{align}
{\cal Y}_{LM_L}^{l_1l_2}(\vec{r}_1,\vec{r}_2) \underset{\scriptscriptstyle l_1+l_2 = L}{=}&\
 \frac{1}{L!}\,\big( \vec{r}_1 \cdot \vec{\nabla}_{\xi}\big)^{l_1}\,
 \big( \vec{r}_2 \cdot \vec{\nabla}_{\xi}\big)^{l_2}\,{\cal Y}_{LM}(\vec{\xi})\,,
\\
{\cal Y}_{LM_L}^{l_1l_2}(\vec{r}_1,\vec{r}_2) \underset{\scriptscriptstyle l_1+l_2 = L+1}{=}&\
\frac{1}{L!}\, \big( \vec{r}_1 \cdot \vec{\nabla}_{\xi}\big)^{l_1-1}\,
\big( \vec{r}_2 \cdot \vec{\nabla}_{\xi}\big)^{l_2-1}
\nonumber \\ & \times
 \big( \vec{R} \cdot \vec{\nabla}_{\xi}\big)\,
 {\cal Y}_{LM}(\vec{\xi})\,,
\end{align}
where $\vec{R} \equiv \vec{r}_1\times \vec{r}_2$,  $\vec{\xi}$ is an arbitrary vector, and the
right-hand-side of the above equations does not depend on $\vec{\xi}$ after the $L$-fold
differentiation.

The bipolar solid harmonics are proportional to the corresponding bipolar spherical harmonics
with a prefactor that does not depend on angles, so their angular parts are exactly the same.
Now, using Eq.~(\ref{sum1}), we obtain that the bipolar solid harmonics ${\cal
Y}_{LM_L}^{l_1l_2}$ obey the analogous summation rule
\begin{align}
\frac{1}{2\,L+1}\sum_{M_L=-L}^L&\
{\cal Y}_{LM_L}^{{l'_1l'_2}^*}(\vec{r}_1^{\,\prime},\vec{r}^{\,\prime}_2)\,{\cal Y}_{LM_L}^{l_1l_2}(\vec{r}_1,\vec{r}_2)
\nonumber \\=&\
{\cal Y}_{i_1..i_L}^{{l'_1l'_2}^*}(\vec{r}^{\,\prime}_1,\vec{r}^{\,\prime}_2)\,{\cal Y}_{i_1 .. i_L}^{l_1l_2}(\vec{r}_1,\vec{r}_2)\,,\label{eq:8}
\end{align}
where ${\cal Y}_{i_1 .. i_L}^{l_1l_2}$ are the symmetric and traceless tensors of rank $L$
with Cartesian indices $i_1 \ldots i_L$,
\begin{align}\label{eq:8b}
{\cal Y}_{i_1 .. i_L}^{l_1l_2}(\vec{r}_1,\vec{r}_2)  \underset{\scriptscriptstyle l_1+l_2 = L}{=} &\
 (r_1^{i_1}\!\!\ldots r_1^{i_{l_1}} r_2^{i_{l_1+1}}\!\!\!\!\ldots r_2^{i_{L}})^{(L)}\,, \\
{\cal Y}_{i_1 .. i_L}^{l_1l_2}(\vec{r}_1,\vec{r}_2)  \underset{\scriptscriptstyle l_1+l_2 = L-1}{=} &\
(r_1^{i_1}\!\!\ldots r_1^{i_{l_1}} r_2^{i_{l_1+1}}\!\!\!\!\ldots r_2^{i_{L-1}} R^{i_L})^{(L)}\,.
\label{eq:8b2}
\end{align}

The summation formula (\ref{eq:8}) shows that the matrix elements with the spatial wave function
\begin{align}
\psi_{ LM_L} =&\  \sum_{l_1l_2} f_{l_1l_2}(r_1,r_2,r)\, {\cal Y}_{LM_L}^{l_1l_2}(\vec{r}_1,\vec{r}_2)
\end{align}
can be represented in terms of matrix elements with the Cartesian wave function
\begin{align}\label{eqq:13}
 \psi^{i_1i_2..i_{L}} = \sum_{l_1l_2} F_{l_1l_2}(r_1,r_2,r)\, {\cal Y}_{i_1i_2 .. i_L}^{l_1l_2}(\vec{r}_1,\vec{r}_2)\,
\end{align}
as follows
\begin{align}
\frac{1}{2\,L+1}\,\sum_{M_L} \langle \psi'_{ LM_L} | Q | \psi_{ LM_L}\rangle = \langle \psi'^{i_1i_2..i_L} | Q | \psi^{i_1i_2..i_L} \rangle\,,
\end{align}
where $Q$ is an arbitrary spatial operator. Eq.~(\ref{eqq:13}) is the Cartesian representation of
the spatial wave function used in the present work.

We now present explicit formulas for the Cartesian wave functions for different values of the
angular momentum and parity. For $L = 0$ we have $l_1 = l_2 = 0$ and only even parity. The wave
function is just a scalar,
\begin{align}\label{eq:9}
\psi\left(^{1,3}\!S^e\right) = F(r_1,r_2,r) \pm (1\leftrightarrow2)\,,
\end{align}
where the upper sign in $\pm$ corresponds to the singlet and the lower sign to the triplet
state. For $L = 1$, we have $(l_1,l_2) = (0,1), (1,0)$ for the odd parity and $(l_1,l_2) = (1,1)$
for the even parity. The corresponding wave functions are vectors,
\begin{align}
\vec{\psi}\left(^{1,3}\!P^o\right) &\ = \vec{r}_1 \, F(r_1,r_2,r) \pm (1\leftrightarrow2)\,, \\
\vec{\psi}\left(^{1,3}\!P^e\right) &\ = \big(\vec{r}_1\times \vec{r}_2\big) \, F(r_1,r_2,r) \pm (1\leftrightarrow2)\,.
\end{align}
The $L = 2$ odd and even wave functions are second-rank tensors,
\begin{align}
\psi^{ij}\left(^{1,3}\!D^o\right) &\ = \bigl[ r_1^i \big(\vec{r}_1\times \vec{r}_2\big)^j + r_1^j  \big(\vec{r}_1\times \vec{r}_2\big)^i\bigr]\, F
\pm (1\leftrightarrow2)\,,
 \\
 \label{6}
\psi^{ij}\left(^{1,3}\!D^e\right) &\ = \big(r_1^ir_1^j\big)^{(2)}\,F + \big(r_1^ir_2^j\big)^{(2)}\,G
\pm (1\leftrightarrow2)\,,
\end{align}
where we suppressed arguments of the radial functions $F$ and $G$ and the elementary second-rank
tensors are defined as
\begin{align}
\big(r_a^ir_b^j\big)^{(2)} &\ = \frac12\Big( r_a^ir_b^j + r_b^ir_a^j -\frac{2}{3}\, \delta^{ij}\, \vec{r}_a\cdot\vec{r}_b\Big)\,\,.
\end{align}
Explicit expressions for the $L = 3$ and $L = 4$ functions can be found in Appendix~A of
Ref.~\cite{wienczek:19}. The spatial wave functions are normalized by
\begin{align}
\big< \psi'(S)|\psi(S)\big> = \big< \psi'^i(P)|\psi^i(P)\big> = \big< \psi'^{ij}(D)|\psi^{ij}(D)\big> = 1\,.
\end{align}

\section{Evaluation of matrix elements}

The spin-dependent wave function with definite values of the total momentum $J$, its projection
$M$, the angular momentum $L$, and the spin $S$ is given by
\begin{align}
\psi_{JM} =&\  \sum_{M_L M_S} \langle L M_L S M_S | J M \rangle \, \psi_{ LM_L}\,\chi_{SM_S} \,,
\end{align}
where $M_S$ is the spin projection, $\chi_{SM_S}$ is the spin function, and $\psi_{ LM_L}$ is the
spatial wave function. As described in the previous section, in our calculations we evaluate all
matrix elements in Cartesian coordinates. The spatial wave function with the angular momentum
$L$ is represented in the form (\ref{eqq:13}); namely, as a traceless tensor of rank $L$
symmetric in all Cartesian indices carried by $\vec r_1$, $\vec r_2$, and $\vec r_1\times \vec
r_2$. In addition, it is assumed that the wave function has a definite symmetry with respect to
$\vec r_1 \leftrightarrow \vec r_2$.

The norm and the expectation value of any spin-independent operator are immediately reduced to the
spatial matrix element,
\begin{align}
\langle \psi_{JM} | Q |\psi_{JM}\rangle =&\ \langle \psi^{i_1 i_2..i_L} | Q |\psi^{i_1 i_2 .. i_L}\rangle\,, \label{norm}
\end{align}
where the summation over Cartesian indices is implicit. This equation is sufficient for
determining the nonrelativistic wave function and the nonrelativistic energy. The relativistic
and QED corrections involve operators depending on the electron spin. The expectation value of an
arbitrary operator $Q$ on a state with definite $J$, for the singlet $S=0$ states, is expressed
as
\begin{align} \label{singlet}
\frac{1}{2J+1}\sum_M & \,\langle \psi_{JM} | Q |\psi_{JM}\rangle
\nonumber  \\
= &\ {\rm Tr} \Bigl[\langle \psi^{i_1 i_2..i_L} | Q |\psi^{i_1 i_2 .. i_L}\rangle\,\Bigl(I-\frac{\vec S{\,}^2}{2}\Bigr) \Bigr]\,,
\end{align}
where $I$ is the unity matrix, $\vec J = \vec L+\vec S$, $\vec S =
(\vec\sigma_1+\vec\sigma_2)/2$, and the trace is performed in the 4-dimensional space of two
spins. Further evaluation of the matrix element proceeds by performing the trace of the operators
in the spin space, with help of the following trace rules,
\begin{eqnarray}
{\rm Tr}\,I &=& 4\,,\\
{\rm Tr}\,S^i &=& 0\,,\\
{\rm Tr}\,S^i\,S^j &=& 2\,\delta^{ij}\,,\\
{\rm Tr}\,S^i\,S^j\,S^k &=& i\,\epsilon^{ijk}\,,\\
{\rm Tr}\,S^i\,S^j\,S^k\,S^l &=&
\delta^{ij}\,\delta^{kl} + \delta^{jk}\,\delta^{il}\,.
\end{eqnarray}
In the case of a spin-independent operator $Q$, Eq.~(\ref{singlet}) is reduced to
Eq.~(\ref{norm}).

For the triplet states one considers three values of $J=L-1,L,L+1$. The expectation value then
takes the form
\begin{align}
&\langle \psi | Q |\psi\rangle_J = \ {\rm Tr} \Bigl[\langle \psi^{j i_2..i_L} | Q |\psi^{i i_2 .. i_L}\rangle\nonumber \\
&\times\Bigl(\frac{\delta^{ij}}{3}\,\vec S^{\,2}\,\Bigl(\frac12-A_{JL}-B_{JL}\Bigr) + S^i\, S^j \,A_{JL} + S^j\, S^i\,B_{JL}\Bigr) \Bigr] \label{triplet}
\,.
\end{align}
For the spin-independent operators, this equation is equivalent to Eq.~(\ref{norm}). The
coefficients $A_{JL}$ and $B_{JL}$ are obtained  by considering two particular cases,  $Q= L^i
S^i$ and $Q=(L^i\,L^j)^{(2)}\,(S^i\,S^j)^{(2)}$. The left-hand-side of Eq.~(\ref{triplet}) is
then immediately expressed in terms of $J$ and $L$, whereas the right-hand-side is evaluated by
using
\begin{align}
-i\,\epsilon^{ijk}\,\langle \psi^{i i_2..i_L} | L^j |\psi^{k i_2 .. i_L}\rangle =&\  (L+1)\,\langle \psi | \psi\rangle\,,\\
\langle \psi^{i i_2..i_L} | (L^i\,L^j)^{(2)} |\psi^{j i_2 .. i_L}\rangle =&\ -\frac{ (L+1)\,(2\,L+3)}{6}\,\langle \psi | \psi \rangle\,.
\end{align}
This consideration gives
\begin{align}
A_{JL} =&\ \frac{L}{L+1}\, \biggl\{-\frac{2\,L+1}{2\,L+3}\,,\,1\,,\,0\biggr\}\,,\label{AJL}\\
B_{JL} =&\ \frac{1}{L+1}\,\biggl\{ \frac{2\,L}{2\,L+3}\,,\, L-1\,,\,-L-1 \biggr\}\,,\label{BJL}
\end{align}
for $J=L+1$, $L$, and $L-1$, correspondingly. These are all the formulas needed to factorize out
the spin dependence of matrix elements and to express them in terms of spatial integrals.

The expectation values of an arbitrary operator $Q$ for the singlet and triplet wave functions
are obtained from Eqs.~(\ref{singlet}) and (\ref{triplet}). We now write explicitly the
corresponding expressions. The results for the $S$ states are
\begin{align} \langle ^1S_0| Q |^1S_0\rangle =&\ {\rm Tr}\biggl[
\langle^1S| Q |^1S\rangle \biggl(I-\frac{S^2}{2}\biggr)\biggr]\,, \\
\langle ^3S_1| Q  |^3S_1\rangle =&\ {\rm Tr}\biggl[
\langle^3S| Q |^3S\rangle\,\frac{S^2}{6}\biggr]\,.
\end{align}
\begin{widetext}
For the $P$ states, we obtain
\begin{align}\label{eqP:1}
\langle {}^1P_1 | Q | {}^1P_1\rangle =&\ {\rm Tr}\biggl[
\langle {}^1P^j| Q  | {}^1P^i\rangle\,\delta^{ij}\biggl(I-\frac{S^2}{2}\biggr)\biggr]\,,\\
\langle {}^3P_0| Q  | {}^3P_0\rangle =&\ {\rm Tr}\biggl[
\langle {}^3P^j| Q  | {}^3P^i\rangle\,
\biggl(\delta^{ij}\,\frac{S^2}{2}-S^j\,S^i\biggr)\biggr]\,, \\
\langle {}^3P_1| Q | {}^3P_1\rangle =&\ {\rm Tr}\biggl[
\langle {}^3P^j| Q | {}^3P^i\rangle\,\frac{1}{2}\,S^i\,S^j\biggr]\,, \\
\langle {}^3P_2| Q | {}^3P_2\rangle =&\ {\rm Tr}\biggl[
\langle {}^3P^j| Q | {}^3P^i\rangle\, \frac{1}{10}
\biggl(2\,S^2\,\delta^{ij}-3\,S^i\,S^j+2\,S^j\,S^i\biggr)\biggr]\,.
\label{eqP:3}
\end{align}
The results for the $D$ states are
\begin{align}
\langle {}^1D_{2}| Q  | {}^1D_{2}\rangle\,=&\ {\rm Tr}\biggl[
\langle {}^1D^{ij}| Q | {}^1D^{ij}\rangle\,\Bigl(I-\frac12\, S^{\,2}\Bigr)\biggr] \,,\\
\langle {}^3D_{1}| Q  | {}^3D_{1}\rangle\,=&\ {\rm Tr}\biggl[
\langle {}^3D^{jk}| Q | {}^3D^{ik}\rangle\,
  \Bigl(\frac12\,\delta^{ij}\,  S^{\,2} -S^j\,S^i\Bigr)\biggr] \,,\\
\langle {}^3D_{2}| Q   | {}^3D_{2}\rangle =&\ {\rm Tr}\biggl[
\langle {}^3D^{jk}| Q   | {}^3D^{ik}\rangle\,
  \Bigl(-\frac{1}{6}\,\delta^{ij}\,  S^{\,2}
  +\frac{2}{3}\,S^i\,S^j +\frac{1}{3}\,S^j\,S^i\Bigr)\biggr] \,,\\
 \langle ^3D_{3}| Q  | {}^3D_{3}\rangle\, =&\ {\rm Tr}\biggl[
\langle ^3D^{jk}| Q  | {}^3D^{ik}\rangle\,
  \Bigl(\frac{11}{42}\,\delta^{ij}\,  S^{\,2}
  -\frac{10}{21}\,\,S^i\,S^j +\frac{4}{21}\,S^j\,S^i\Bigr)\biggr]\,.
  \label{tr2}
  \end{align}
\end{widetext}

\section{Integrals with exponential basis functions}
The radial parts of the wave function (\ref{eqq:13}) are represented as linear combinations of
the exponential basis functions,
\begin{align}\label{eq2:11}
F(r_1,r_2,r) = \sum_{k=1}^N c_k\,e^{-\alpha_k r_1-\beta_k r_2-\gamma_k r}\,,
\end{align}
where $c_k$ are linear coefficients, $N$ is the size of the basis, and $\alpha_k$, $\beta_k$, and
$\gamma_k$ are nonlinear parameters obtained in the process of the basis optimization.
One of the great features of the exponential basis functions is that the evaluation of radial
integrals is very simple. A calculation of radial matrix elements of various operators with wave
functions of the form (\ref{eq2:11}) is reduced to evaluation of the integrals $I(i,j,k)$,
\begin{equation}
I(i,j,k) =
\frac{1}{16\pi^2}\!\int d^3r_1\!\int d^3r_2\, r_1^{i-1}\,r_2^{j-1}\,r^{k-1} e^{-\alpha r_1-\beta r_2-\gamma r}
\,.
\end{equation}
For matrix elements of the nonrelativistic Hamiltonian, only integrals with non-negative values
of $i$, $j$, and $k$ are required. All such integrals can be obtained by differentiation of the
master integral $I(0,0,0)$ over the nonlinear parameters,
\begin{equation}\label{eq2:12c}
I(n_i,n_j,n_k) = (-1)^{n_i+n_j+n_k}\,\frac{\partial^{n_i}}{\partial\alpha^{n_i}}
\,\frac{\partial^{n_j}}{\partial\beta^{n_j}}
\,\frac{\partial^{n_k}}{\partial\gamma^{n_k}}
\,I(0,0,0)\,,
\end{equation}
for $n_i,n_j,n_k \ge 0$. The expression for the master integral $I(0,0,0)$ is very simple,
\begin{equation}\label{eq2:13}
I(0,0,0) = \frac{1}{(\alpha+\beta)(\beta+\gamma)(\gamma+\alpha)}\,.
\end{equation}

Matrix elements of relativistic corrections involve integrals with additional inverse powers of
$r_1$, $r_2$, and $r$, whose evaluation requires two additional master integrals. Their
expression can be obtained by integrating Eq.~(\ref{eq2:13}) with respect to the corresponding
nonlinear parameters. The results are
\begin{align}\label{eq2:14}
I(0,0,-1)=&\
\frac{1}{(\alpha+\beta)(\alpha-\beta)}\,\ln\bigg(\frac{\alpha+\gamma}{\beta+\gamma}\bigg) \,, \\
I(-1,0,-1)=&\
\frac{1}{2\,\beta}\bigg[\frac{\pi^2}{6}+\frac{1}{2}\,\ln^2\bigg(\frac{\alpha+\beta}{\beta+\gamma}\bigg)
 \nonumber \\ &
+{\rm Li}_2\bigg(1-\frac{\alpha+\gamma}{\alpha+\beta}\bigg) + {\rm
Li}_2\bigg(1-\frac{\alpha+\gamma}{\beta+\gamma}\bigg) \bigg] \,,
\end{align}
where ${\rm Li}_2$ is the dilogarithm function \cite{lewin:book}.  Other integrals for
relativistic corrections are obtained by differentiating the above formulas for master integrals.

We note that Eq.~(\ref{eq2:14}) contains a spurious singularity at $\alpha = \beta$. The zero
in the denominator is compensated by the vanishing logarithm function and thus is not a real
singularity but can lead to numerical instabilities. In order to transform Eq.~(\ref{eq2:14}) to
an explicitly regular form, we introduce a regularized logarithm function $\overline{\ln}_1(x)$
by separating out the first term of the Taylor expansion,
\begin{align}
\ln(1+x) \equiv x\,\overline{\ln}_1(x)\,.
\end{align}
Introducing $\overline{\ln}_1(x)$ with $x = (\alpha-\beta)/(\beta+\gamma)$ in Eq.~(\ref{eq2:14}),
we obtain a regular representation of this formula. In practical calculations we encounter more
spurious singularities of this kind. They are eliminated with the help of functions
$\overline{\ln}_n(x)$, which are introduced analogously to $\overline{\ln}_1(x)$ by separating
$n$ first terms of the Taylor expansion of $\ln(1+x)$.

Matrix elements of QED corrections involve several integrals with large negative powers of radial
distances, like $1/r^3$, $1/r^4$, and even $1/r^5$. Such integrals are singular and need proper
definitions. With the exponential functions, it is possible to obtain simple and numerically
stable representations for such integrals. The corresponding procedure is described in
Appendix~\ref{app:singularintegrals}. Numerical results for basic singular integrals for the
$2^3\!S$ and $2^3\!P$ states of helium are presented in Table~\ref{tab:oprsQ1}.

\begin{table}
\caption{Expectation values of singular operators for the $2^3\!S$ and $2^3\!P$ states
of helium, in atomic units. The numerical uncertainty is less than the last significant digit.
\label{tab:oprsQ1}}
\begin{ruledtabular}
\begin{tabular}{c w{5.10}w{5.10}}
 & \multicolumn{1}{c}{$2^3S$}
 & \multicolumn{1}{c}{$2^3P$}
  \\ \hline
 $1/r^3$                   	                     &  0.038\,861  &  0.047\,927 \\
 $1/r^4$                  	                     &  0.026\,567  &  0.043\,348 \\
 $1/r^5$               	                         &  0.017\,580  &  0.027\,240 \\
 $1/r_1^3$   	  	                             &-23.022\,535  &-21.886\,142 \\
 $1/r_1^4$   			                         & 25.511\,837  & 24.525\,751 \\
\end{tabular}
\end{ruledtabular}
\end{table}

In our calculations of the $\alpha^7\,m$ QED effects \cite{patkos:21:helamb}, integrals with $\ln
r$ were encountered for the first time,
\begin{align}
I_{\rm log}(i,j,k) = &\
\frac{1}{16\pi^2}\!\int d^3r_1\!\int d^3r_2\, r_1^{i-1}\,r_2^{j-1}\,r^{k-1}
 \nonumber \\  & \times
 \,(\ln r+\gamma_E)
 \,e^{-\alpha r_1-\beta r_2-\gamma r}
\,,
\end{align}
where $\gamma_E$ is the Euler gamma constant. Such integrals are evaluated with the help of the
following set of master integrals \cite{patkos:21:helamb}:
\begin{align}
I_{\rm log}(0,0,0) =&\
\frac{1}{(\alpha-\beta)\,(\alpha+\beta)}\biggl[\frac{\ln(\alpha+\gamma)}{\alpha+\gamma} - \frac{\ln(\beta+\gamma)}{\beta+\gamma}\biggr]\,,\\
I_{\rm log}(0,0,-1) =&\
\frac{1}{2\,(\alpha-\beta)\,(\alpha+\beta)}\bigl[\ln^2(\beta+\gamma) - \ln^2(\alpha+\gamma)\bigr]\,,
\end{align}
\begin{widetext}
\begin{align}\label{eq2:20}
I_{\rm log}(-1,0,-1) \underset{\alpha>\beta}{=} &\
\frac{1}{2\,\beta}\bigg\{
\frac{1}{2}\,\ln\bigg(\frac{\alpha-\beta}{\alpha+\beta}\bigg)\,\big[\ln^2(\alpha+\gamma) - \ln^2(\beta+\gamma)\big]
 \nonumber \\ &
+ \ln(\alpha+\gamma)\bigg[{\rm Li}_2\bigg(\frac{-\beta + \gamma}{\alpha + \gamma}\bigg) - {\rm Li}_2\bigg(\frac{\beta + \gamma}{\alpha + \gamma}\bigg)\biggr]
+ {\rm Li}_3\bigg(\frac{-\beta + \gamma}{\alpha + \gamma}\bigg) - {\rm Li}_3\bigg(\frac{\beta + \gamma}{\alpha + \gamma}\bigg)\bigg\}\,,
\end{align}
\end{widetext}
where ${\rm Li}_3$ is the trilogarithm function \cite{lewin:book}. Eq.~(\ref{eq2:20}) is valid
for $\alpha>\beta$. The corresponding result for $\alpha<\beta$ is obtained by the analytic
continuation with help of the following identities \cite{lewin:book}
\begin{align}
{\rm Li}_2\big(-z\big) + {\rm Li}_2\big(-z^{-1}\big)  =&\  -\frac{\pi^2}{6}-\frac{\ln^2(z)}{2}\,, \\
{\rm Li}_3\big(-z\big) - {\rm Li}_3\big(-z^{-1}\big) = &\ -\frac{\pi^2}{6} \ln(z) -\frac{1}{6}\,\ln^3(z)\,.
\end{align}
The result for the case of $\alpha = \beta$ is straightforwardly obtained from
Eq.~(\ref{eq2:20}).

\section{Nonrelativistic energy and wave function}

\begin{table*}
\caption{Convergence study of the nonrelativistic energy $E_0$ of the $2\,^3\!P$ state of He, for the
infinitely heavy nucleus, in atomic units. $N$ is the size of the basis.
\label{tab:NRenergy}}
\begin{tabular}{c w{10.35}w{4.9}}
\hline
 \multicolumn{1}{c}{$N$}
 & \multicolumn{1}{c}{$E_0$}
         & \multicolumn{1}{c}{Increment}
                        \\\hline\\[-7pt]
  100  &  -2.133\,164\,189\,889\,061\,228\,337\,6   \\
  200  &  -2.133\,164\,190\,766\,840\,570\,131\,8   & -0.89\times 10^{-9}  \\
  400  &  -2.133\,164\,190\,779\,088\,013\,045\,2   & -0.12\times 10^{-10}  \\
  800  &  -2.133\,164\,190\,779\,281\,832\,163\,4   & -0.20\times 10^{-12}  \\
 1200  &  -2.133\,164\,190\,779\,283\,169\,438\,0   & -0.14\times 10^{-14}  \\
 1600  &  -2.133\,164\,190\,779\,283\,201\,696\,6   & -0.33\times 10^{-16}  \\
 2000  &  -2.133\,164\,190\,779\,283\,204\,908\,9   & -0.32\times 10^{-17}  \\
 2400  &  -2.133\,164\,190\,779\,283\,205\,102\,6   & -0.20\times 10^{-18}  \\
 2800  &  -2.133\,164\,190\,779\,283\,205\,142\,0   & -0.39\times 10^{-19}  \\
 3200  &  -2.133\,164\,190\,779\,283\,205\,145\,6   & -0.36\times 10^{-20}  \\
 3600  &  -2.133\,164\,190\,779\,283\,205\,146\,4   & -0.87\times 10^{-21}  \\
Ref.~\cite{aznabaev:18}
       &  -2.133\,164\,190\,779\,283\,205\,146\,992\,763\,806\\
\hline
\end{tabular}
\end{table*}

The nonrelativistic Hamiltonian of the helium atom for the infinitely heavy nucleus is
\begin{equation}\label{eq:24}
H_0 = \frac{\vec{p}_1^{\,\,2}}{2}+ \frac{\vec{p}_2^{\,\,2}}{2} - \frac{Z}{r_1} - \frac{Z}{r_2} + \frac{1}{r}\,,
\end{equation}
where $\vec{p}_a = -i\vec{\nabla}_a$ is the momentum operator of the electron $a$ and $Z$ is the
nuclear charge number ($Z = 2$ for helium). The Schr\"odinger equation is
\begin{equation}
H_0\, \psi(\bfr_1,\bfr_2) = E_0 \, \psi(\bfr_1,\bfr_2)\,.
\end{equation}
A direct solution of the Schr\"odinger equation is standardly substituted by the problem of
finding the minimum or, generally, a stationary point of the variational functional
\begin{equation}
\Phi(\psi) = \frac{ \lbr \psi \big| H_0 \big| \psi \rbr }{\lbr \psi| \psi\rbr}\,.
\end{equation}
The variational eigenvalues obtained from this functional are the upper bounds to the true
eigenvalues, and the corresponding eigenvectors provide the linear coefficients $c_k$ of the
basis-set expansion (\ref{eq2:11}). It is important that the variational principle works equally
well for the ground and for the excited states.

The finite nuclear mass correction to the nonrelativistic energy is induced by the nuclear
kinetic energy operator
\begin{equation} \label{eq:27}
\delta_M H = \frac{\vec{P}^{\,2}}{2M}\,,
\end{equation}
where $M$ is the nuclear mass and $\vec{P} = -\vec p_1-\vec p_2$ is the nuclear momentum. There
are two ways to incorporate the nuclear mass effect to the nonrelativistic energy: (i) to include
the operator $\delta_{M} H $ into the nonrelativistic Hamiltonian $H_0$ and solve the
nuclear-mass dependent Schr\"odinger equation and (ii) to solve the Schr\"odinger equation for
the infinitely heavy nucleus and to account for the nuclear mass effects by perturbation theory.

In our calculations with the exponential basis we found that the inclusion of $\delta_{M}H$ into
the nonrelativistic Hamiltonian leads to numerical instabilities for $S$ states (but not for $P$
and higher-$L$ states). So, for $S$ states we account for the nuclear mass effects by
perturbation theory (up to the third order in $1/M$ \cite{pachucki:17:heSummary}), whereas for the
$P$ and $D$ states we usually include $\delta_{M} H$ into the solution of the Schr\"odinger
equation. We checked that for the $P$ and $D$ states both methods yield equivalent results.

It should be mentioned that in the literature it is customary to split the operator $\delta_M H$ into
the mass-scaling and mass-polarization parts,
\begin{equation} \label{eq:28}
\delta_M H = \frac{\vec{p}_1^{\,\,2}+\vec{p}_1^{\,\,2}}{2M} +
\frac{\vec{p}_1\cdot\vec{p}_2}{M}\,.
\end{equation}
The effect of the mass scaling (caused by the first term in Eq.~(\ref{eq:28})) can be
incorporated into the nonrelativistic Hamiltonian (\ref{eq:24}) by switching to the reduced-mass
atomic units $r\to \mu\,r$, where $\mu = 1/(1+m/M)$ is the reduced mass. As a result, the
mass-scaling term leads to the appearance of the reduced mass prefactor in the nonrelativistic
energy $E_0 \to \mu\,E_0$ and only the mass polarization term needs to be accounted for
separately. We find it more convenient to keep the nuclear kinetic energy operator in the closed
form (\ref{eq:27}), because this greatly simplifies consideration of higher-order recoil QED
effects.

Because the nonrelativistic Hamiltonian $H_0$ does not depend on spin, its matrix elements are
immediately reduced to radial integrals with the spatial wave functions according to
Eq.~(\ref{norm}). Computing the action of gradients $\nabla_{1,2}$ on the wave functions
(\ref{eqq:13}), we express the matrix elements $\lbr \psi| H_0|\psi\rbr$ as a linear combination
of integrals $I(i,j,k)$ with $i,j,k \ge 0$, which are rational functions of the nonlinear
parameters $\alpha_n$, $\beta_n$, and $\gamma_n$.

The choice of the nonlinear basis parameters $\alpha_n$, $\beta_n$, and $\gamma_n$ is crucially
important for obtaining an accurate and compact representation of the wave function and the
energy $E_0$. The general approach is to perform the variational optimization of the basis
parameters, by searching for a minimum of the eigenvalue of the Hamiltonian matrix corresponding
to the desired reference state. Because the optimization of each individual nonlinear parameter is
not effective from the computational point of view, we use the approach introduced by Vladimir
Korobov \cite{korobov:00}. In this method, the (real) nonlinear parameters $\alpha$, $\beta$,
and $\gamma$ are quasirandomly distributed in the intervals
\begin{eqnarray} \label{eq:29}
\alpha &\in& [A_1,A_2]\,, \nonumber \\
\beta &\in& [B_1,B_2]\,, \nonumber \\
\gamma &\in& [C_1,C_2]\,,
\end{eqnarray}
and the parameters $A_{1,2}$, $B_{1,2}$, and $C_{1,2}$ are determined by the variational
optimization. We note that the nonlinear parameters as well as $A_{1,2}$, $B_{1,2}$, and
$C_{1,2}$ can be both positive and negative. However, in order to ensure the normalizability of
the wave function and its physical behavior at large $r_1$, $r_2$, and $r$, we require that
\begin{eqnarray}
\left\{ \alpha+\beta, \alpha+\gamma, \beta+\gamma\right\} >\epsilon \,,
\end{eqnarray}
where $\epsilon\sim\sqrt{2\,E_{\rm io}}$, with $E_{\rm io}$ being the ionization energy.
The performance of the basis set can be significantly improved if one introduces several sets of
intervals $A_{1,2}$, $B_{1,2}$, and $C_{1,2}$ which are optimised variationally. In our
calculations we use typically two or three sets of intervals. This can be considered as an
analogue of several different exponential scales in the Hylleraas-type calculations by Drake {\em at al.}
 \cite{drake:92,drake:01}.

We also note that in calculations for excited $1snl$ states it is advantageous to include
several screened hydrogenic wave functions of the type
$\phi^Z_{1s}(\vec{r}_1)\,\phi^{Z-1}_{n'l}(\vec{r}_2)$ with $n' \le n$ in the
basis, whose parameters are
excluded from optimization. This ensures that the variational optimization is localized at the
local minimum with the desired principal quantum number $n$ and does not collapse to lower $n$'s.

Our procedure for determination of the nonrelativistic wave function and energy looks as follows.
For a given size of the basis $N$ the nonlinear parameters $\alpha_n$, $\beta_n$, and $\gamma_n$
with $n=1,\ldots ,N$ are distributed quasirandomly within the initial set of intervals with
parameters $A_i$, $B_i$, and $C_i$. Then, the $N\times N$ matrix of the nonrelativistic
Hamiltonian $H_0$ is computed. The linear coefficients $c_n$ and the desired reference-state
eigenvalue $E_0$ are determined by the inverse iteration method. The inversion of the Hamiltonian
matrix is performed by the LDU decomposition method. This procedure is repeated for different
sets of the parameters $A_i$, $B_i$, and $C_i$, searching for the minimum value of the energy
eigenvalue.

A disadvantage of working with the exponential basis is that the basis quickly degenerates as $N$
is increased (i.e. the determinant of the Hamiltonian matrix becomes very small), which leads to
numerical instabilities in linear algebra routines. Because of this the usage of an
extended-precision arithmetics is mandatory. In our calculations we used the Fortran~95 libraries
for the octuple-precision (about 64 digits) arithmetics written by V.~Korobov
\cite{korobov:priv}, quad-double routine by D.~H.~Bailey, and  MPFUN/MPFR library by D.~H.~Bailey
\cite{bailey:mpfun}.

Table~\ref{tab:NRenergy} shows an example of the convergence of numerical results with the
exponential basis with increase of the basis size. We observe that with just $N = 200$ basis
functions one obtains the nonrelativistic energy with about 10-digit accuracy.

\section{Relativistic correction}

The relativistic correction splits the nonrelativistic energy levels with quantum numbers $L > 0$
and $S>0$ into sublevels according to the value of the total momentum $J$. This effect is known
as the fine structure. It is often convenient to consider separately the centroid energy levels
obtained by averaging over all $J$ sublevels, and the fine-structure intervals between individual
$J$ sublevels. The centroid energy is defined as
\begin{eqnarray}
E(^nL) &=& \frac{\sum_{JM} E(^nL_{JM})}{(2L+1)(2S+1)} = \frac{\sum_{J}(2J+1) E(^nL_J)}{(2L+1)(2S+1)}
\,. \nonumber \\
\end{eqnarray}

The relativistic correction is induced by the Breit Hamiltonian, which is conveniently separated
into the spin-independent and the spin-dependent parts,
\begin{eqnarray}
H_{\rm Breit} = H_A + H_{\rm fs}\,.
\end{eqnarray}
In the leading order of perturbation theory, the spin-independent part $H_A$ contributes only to
the centroid energy, whereas the spin-dependent part $H_{\rm fs}$ causes the fine structure
splitting.

\subsection{Centroid energy}

The spin-independent part of the Breit Hamiltonian is given by
\begin{eqnarray}\label{eq:33}
H_{A}&=& -\frac{1}{8}\,(\vec{p}_1^{\,\,4}+\vec{p}_2^{\,\,4})+
\frac{Z\,\pi}{2}\,\big[\delta^3(r_1)+\delta^3(r_2)\big]
\nonumber \\ &&
+\pi\,\delta^3(r)
-\frac{1}{2}\,p_1^i\,
\biggl(\frac{\delta^{ij}}{r}+\frac{r^i\,r^j}{r^3}\biggr)\,p_2^j\
\nonumber \\ &&
+
 \frac{Z}{2M}\,\biggl[
p_1^i\,\biggl(\frac{\delta^{ij}}{r_1} + \frac{r_1^i\,r_1^j}{r_1^3}\biggr)+
p_2^i\,\biggl(\frac{\delta^{ij}}{r_2} + \frac{r_2^i\,r_2^j}{r_2^3}\biggr)\biggr]\,P^j\,,\nonumber\\
\end{eqnarray}
where $\vec{P} = -\vec p_1-\vec p_2$ is the nuclear momentum. In order to account for the finite
nuclear mass effects, the expectation value of the operator $H_{A}$ should be evaluated with the
eigenfunctions $\psi_M$ of the Schr\"odinger Hamiltonian with the finite nuclear mass (i.e. the
sum of Eqs.~(\ref{eq:24}) and (\ref{eq:27})). Alternatively, the wave function $\psi_M$ can be
constructed by perturbation theory in $1/M$. In our calculations, we include the nuclear recoil
effect for the relativistic correction perturbatively for the $S$ states, and nonperturbatively
for the $L > 0$ states.

The matrix element of $H_{A}$ is reduced to the radial integral with the spatial wave functions
according to Eq.~(\ref{norm}) and can be evaluated numerically. However, the expectation values
of the operators $\vec{p}_a^{\,\,4}$ and $\delta^3(r_a)$ are slowly converging with respect to
the size of the basis because these operators are nearly singular. It is possible to significantly
improve the speed of convergence if one transforms these operators to a more regular form
\cite{drachman:81}. Specifically, for a given nearly singular operator $H_X$ we search for
another, more regular operator $H_{XR}$ and an additional operator $Q_X$, which satisfy the
following equation
\begin{align}\label{eq:34}
H_X = H_{XR} + \big\{ H_0-E_0, Q_X\big\}\,,
\end{align}
where $\{.\,,.\}$ denotes the anticommutator. It is obvious that $\lbr H_X \rbr = \lbr H_{XR}
\rbr$, as long as the expectation value is evaluated with the eigenfunctions of the Hamiltonian
$H_0$. In practice, it is usually possible to find such a pair of operators $H_{XR}, Q_X$ that
the most singular part of $H_X$ is absorbed in the anticommutator. The additional operator $Q_X$
is generally a combination of $Z/r_1$, $Z/r_2$, and $1/r$, with the coefficients in front of
these terms determined by requiring the cancellation of all Dirac-$\delta$-like contributions.

Specifically, we find the following regularized form of the operator $H_A$ (without the nuclear
recoil) \cite{pachucki:06:hesinglet}
\begin{align} \label{eq:35}
H_{AR} =&\ -\frac12\big( E_0-V\big)^2
- p_1^i \frac1{2r}\Big(\delta^{ij}+\frac{r^ir^j}{r^2}\Big)p_2^j
 \nonumber \\ &
+ \frac14 \vec{\nabla}_1^2  \vec{\nabla}_2^2 -\frac{Z}{4}\Big( {\frac{\vec{r}_1}{r_1^3}\cdot
\vec{\nabla}_1 + \frac{\vec{r}_2}{r_2^3}\cdot \vec{\nabla}_2} \Big) \,,
\end{align}
where $V = -Z/r_1-Z/r_2+1/r$. The operator $\vec{\nabla}_1^2  \vec{\nabla}_2^2$ in the above
formula is not self-adjoint and requires an explicit definition. Its action on a trial function
$\phi$ on the right should be understood as plain differentiation (omitting $\delta^3(r)$); no
differentiation by parts is allowed in the matrix element. It can be checked that the operators
$H_A$ and $H_{AR}$ satisfy the following equation
\begin{align}
H_A = H_{AR} + \big\{H_0 -E_0,Q\big\}\,,
\end{align}
where
\begin{align}
Q = -\frac14\Big( \frac{Z}{r_1}+\frac{Z}{r_2} - \frac2{r}\Big) \,.
\end{align}
Formulas with the finite nuclear mass are analogous but more lengthy; they are given by
Eqs.~(62)-(67) of Ref.~\cite{patkos:16:triplet}.

Table~\ref{tab:Breit} presents numerical results for the leading relativistic correction to the
$2\,^3\!P$ centroid energy, performed with different basis sets. We observe that, for the same
basis size, the number of correct digits for the matrix element is half as much as for the
nonrelativistic energy.

\begin{table}
\caption{Convergence study of the leading relativistic correction, $\lbr H_{A}\rbr$, for the $2\,^3\!P$ state of He, for the
infinitely heavy nucleus. Units are $\alpha^4m = \alpha^2$\,a.u.
\label{tab:Breit}}
\begin{tabular}{c w{5.15}w{4.9}}
\hline
 \multicolumn{1}{c}{$N$}
 & \multicolumn{1}{c}{$\Delta E$}
         & \multicolumn{1}{c}{Increment}
                        \\\hline\\[-7pt]
 100  &   -1.967\,366\,535\,960  & \\
 200  &   -1.967\,360\,971\,947  &  0.55\times 10^{-5} \\
 400  &   -1.967\,358\,035\,372  &  0.29\times 10^{-5} \\
 600  &   -1.967\,358\,371\,368  & -0.33\times 10^{-6} \\
 800  &   -1.967\,358\,354\,920  &  0.16\times 10^{-7} \\
1200  &   -1.967\,358\,362\,599  & -0.76\times 10^{-8} \\
1600  &   -1.967\,358\,376\,018  & -0.12\times 10^{-7} \\
2000  &   -1.967\,358\,374\,001  &  0.20\times 10^{-8} \\
2400  &   -1.967\,358\,374\,197  & -0.20\times 10^{-9} \\
2800  &   -1.967\,358\,374\,236  & -0.38\times 10^{-10} \\
3200  &   -1.967\,358\,374\,256  & -0.26\times 10^{-10} \\
3600  &   -1.967\,358\,374\,254  &  0.15\times 10^{-11} \\
\hline
\end{tabular}
\end{table}

\subsection{Fine structure}

The fine structure of energy levels is induced by spin-dependent operators. The spin-dependent
part of the Breit Hamiltonian is conveniently written as a sum of three operators with different
spin structure,
\begin{align}
H_{\rm fs} = H_B + H_C + H_D\,,
\end{align}
with
\begin{align}\label{eq:36}
H_B = &\ \bigg[
\frac{Z}{4}\biggl(\frac{\vec{ r}_1}{r_1^3}\times\vec{ p}_1+\frac{\vec{ r}_2}{r_2^3}\times\vec{ p}_2\biggr)\,(1+2\,\kappa)
 \nonumber \\ &
 -\frac{3}{4}\,\frac{\vec{ r}}{r^3}\times(\vec{ p}_1-\vec{
p}_2)\,\biggl(1+\frac{4\,\kappa}{3}\biggr)
 \nonumber \\ &
-
\frac{Z}{2M}\biggl(\frac{\vec{ r}_1}{r_1^3}+\frac{\vec{ r}_2}{r_2^3}\biggr) \times \vec{P}\,(1+\kappa)
\bigg]\,\frac{\vec{\sigma}_1+\vec{\sigma}_2}{2}
 \,,\\
H_C = & \bigg[
\frac{Z}{4}\biggl(\frac{\vec{ r}_1}{r_1^3}\times\vec{ p}_1-\frac{\vec{ r}_2}{r_2^3}\times\vec{ p}_2\biggr)\,(1+2\,\kappa)
 \nonumber \\ &
+\frac{1}{4}\,\frac{\vec{ r}}{r^3}\times
(\vec{ p}_1+\vec{ p}_2)
 \nonumber \\ &
-
\frac{Z}{2M}\biggl(\frac{\vec{ r}_1}{r_1^3}-\frac{\vec{ r}_2}{r_2^3}\biggr) \times \vec{P}\,(1+\kappa)
\bigg]\,\frac{\vec{\sigma}_1-\vec{\sigma}_2}{2}
\,,\\
H_D   = & \frac{1}{4}\left(
\frac{\vec{\sigma}_1\,\vec{\sigma}_2}{r^3}
-3\,\frac{\vec{\sigma}_1\cdot\vec{r}\,\vec{\sigma}_2\cdot\vec{r}}{r^5}\right)\,(1+\kappa)^2
\,,
\label{eq:38}
\end{align}
where $\kappa=\nicefrac{\alpha}{2\pi} + O(\alpha^2)$ is the anomalous magnetic moment correction
and $\vec{\sigma}_a$ is the vector of Pauli matrices acting on $a$'th electron. We note that the
operators $H_B$, $H_C$, and $H_D$ contain radiative corrections in form of the electron anomalous
magnetic moment. In this way we account for the complete QED effects of order $\alpha^5\,m$ to
the fine structure.

It should be mentioned that the matrix element of $H_C$ is nonzero only if the operator is
sandwiched between wave functions with different spin values. Therefore, any symmetrical matrix
element of $H_C$ vanishes, and this operator does not contribute in the leading order of
perturbation theory. We note, however, that $H_C$ contributes to the second-order perturbation
corrections (in the order $\alpha^6\,m$).

In order to perform the spin-angular reduction in the matrix elements of $H_{\rm fs}$, it is
convenient to introduce spatial operators $Q_B$, $Q_C$, and $Q_D$, explicitly separating the
spatial and the spin degrees of freedom,
\begin{eqnarray}
H_B & = &  \vec Q_B \cdot \frac{\vec{\sigma}_1+\vec{\sigma}_2}{2}\,,\\
H_C& = & \vec Q_C \cdot\frac{\vec{\sigma}_1-\vec{\sigma}_2}{2} \,,\label{50}\\
H_D  & = & Q^{ij}_D \,\frac{1}{2}\,\sigma_1^i\,\sigma_2^j
\,.
\end{eqnarray}

Using Eqs.~(\ref{triplet}), (\ref{eqP:1})--(\ref{eqP:3}) and performing traces of the spin
operators, we express all matrix elements in terms of spatial radial integrals. For the $^3P$
states, we obtain
\begin{align}
\frac1{2J+1}\sum_M \lbr{}^3P_{JM}| H_B | {}^3P_{JM} \rbr
  &= i\epsilon_{jkl}\, \lbr ^3P^j | Q_B^k |  ^3P^l\rbr\, u_J(P)\,,
  \\
\frac1{2J+1}\sum_M \lbr{}^3P_{JM}| H_D | {}^3P_{JM} \rbr
  &= \lbr ^3P^j | Q_D^{jl} |  ^3P^l\rbr\, v_J(P)\,,
\end{align}
where
\begin{align}
u_J(P) = &\ (1,\nicefrac12,-\nicefrac12)\,, \\
v_J(P) = &\ (-1,\nicefrac12,-\nicefrac1{10})\,,
\end{align}
for $J = 0$, 1, and 2, respectively.

For the $^3D$ states, an analogous calculation yields
\begin{align}
\frac1{2J+1}\sum_M &\ \lbr{}^3D_{JM}| H_B | {}^3D_{JM} \rbr
 \nonumber \\ &
  = i\epsilon_{jli}\, \lbr ^3D^{jk} | Q_B^l |  ^3D^{ik}\rbr\, u_J(D)\,,
  \\
\frac1{2J+1}\sum_M &\ \lbr{}^3D_{JM}| H_D | {}^3D_{JM} \rbr
 \nonumber \\ &
  = \lbr ^3D^{ik} | Q_D^{ij} |  ^3D^{jk}\rbr\, v_J(D)\,,
\end{align}
where
\begin{align}
u_J(D) = &\ (\nicefrac12,\nicefrac16,-\nicefrac13)\,, \\
v_J(D) = &\ (-1,1,-\nicefrac2{7})\,,
\end{align}
for $J = 1$, 2, and 3, respectively.

\section{Leading QED correction}

The leading QED contribution is of the order $\alpha^5\,m$. For the fine structure, this
contribution is already accounted for by the electron anomalous magnetic moment terms in the
Breit Hamiltonian, as given by Eqs.~(\ref{eq:36})-(\ref{eq:38}). So, we need to examine only the
centroid energy.

The spin-independent $m\alpha^5$ Hamiltonian representing the leading QED effects was derived
in the 1950s by Araki and Sucher \cite{araki:57,sucher:57}
\begin{eqnarray}
H^{(5)} &=&
\left(\frac{19}{30}+\ln(\alpha^{-2}) - \bl \right)\,
\frac{4\,Z}{3}\,\big[ \delta^3(r_1) + \delta^3(r_2)\big]
 \nonumber \\ &&
+ \left(\frac{164}{15}+\frac{14}{3}\,\ln\alpha \right)\,\delta^3(r)
-\frac{7}{6\,\pi}\,\frac{1}{r^3_{\!\epsilon}}
\,,
\end{eqnarray}
where $\bl$ is the so-called Bethe logarithm defined as
\begin{align}\label{eq:55}
\bl &=&
\frac{\big \langle (\vec p_1 + \vec p_2) \,(H_0-E_0)\,\ln \big[2\,(H_0-E_0)\big]\,
(\vec p_1 + \vec p_2) \big\rangle}{2\,\pi\,Z\,\big\langle\delta^3(r_1) + \delta^3(r_2)\big\rangle}\,,
\end{align}
 and $1/r^3_{\!\epsilon}$ is the regularized $1/r^3$ operator (distribution) defined by its matrix
 elements with an arbitrary smooth function $f(\vec{r})$ as
\begin{eqnarray}\label{eq:56}
\int d^3 r\, f(\vec{r})\, \frac{1}{r^3_{\!\epsilon}} =
\lim_{\epsilon\rightarrow 0}\int d^3 r\,
f(\vec{r}) && \biggl[\frac{1}{r^3}\,\Theta(r-\epsilon)
  \nonumber \\ &&
+ 4\,\pi\,\delta^3(r)\,
(\gamma_E+\ln \epsilon)\biggr]\,.
\nonumber \\
\end{eqnarray}

The nuclear recoil correction to the leading QED contribution consists of two parts,
\begin{eqnarray}
E_M^{(5)} &=& 2\,\langle H^{(5)}\,\frac{1}{(E_0-H_0)'}\,\delta_M H\rangle
+ \langle\delta_M H^{(5)}\rangle\,,
\end{eqnarray}
where $\delta_M H$ is defined by Eq.~(\ref{eq:27}), and $\delta_M H^{(5)}$ is the recoil addition
to the $\alpha^5\,m$ Hamiltonian given by \cite{pachucki:00:herec}
\begin{eqnarray}
\delta_M H^{(5)} &=& \frac1M \bigg[
\biggl(\frac{62}{3}+\ln(\alpha^{-2})
-8\,\bl - \frac{4}{Z}\,\delta_M\bl \biggr)\,
 \nonumber \\ && \times
\frac{Z^2}{3}\,\big[ \delta^3(r_1) + \delta^3(r_2)\big]
-\frac{7\,Z^2}{6\,\pi}\,\Big(\frac{1}{r_{1, \epsilon}^3} + \frac{1}{r_{2, \epsilon}^3}
\Big)\bigg]\,.
\nonumber \\
\end{eqnarray}
Here, $\delta_M \ln k_0$ is the correction to the Bethe logarithm $\ln k_0$ induced by the
nonrelativistic kinetic energy operator $\vec{P}^{\,2}/2$, and $1/r^3_{a,\epsilon}$ is the
regularized $1/r^3_{a}$ operator defined analogously to Eq.~(\ref{eq:56}).

The recoil correction to the Bethe logarithm $\delta_M \ln k_0$  is often separated into the
mass-scaling and mass-polarization parts,
\begin{align}
  \delta_M\bl =&\ 1+ \delta_{p_1p_2}\!\bl \,,
\end{align}
where $\delta_{p_1p_2}$ denotes the perturbation due to the mass polarization operator $\vec
p_1\cdot\vec p_2$. The corresponding separation for the $1/r^3_{\!\epsilon}$ matrix element reads
\begin{align}
  \delta_M \Big<\frac{1}{r^3_{\!\epsilon}}\Big> =&\
-3\,\Big<\frac{1}{r^3_{\!\epsilon}}\Big> + \big< 4\,\pi\delta^3(r) \big>
+ \delta_{p_1p_2} \Big<\frac{1}{r^3_{\!\epsilon}}\Big>\,.
\end{align}

From the computational point of view, the numerical evaluation of the QED effects involves two
new features, as compared to the relativistic correction: matrix elements of the singular
operators $1/r^3$ and $1/r_a^3$ and the Bethe logarithm. Calculation of expectation values of
singular operators with exponential basis functions is examined in
Appendix~\ref{app:singularintegrals}; it does not present any computational difficulties. On the
contrary, the computation of the Bethe logarithm is rather nontrivial; it is examined in the next
section.

\subsection{Bethe logarithm}

There are two different approaches developed for the calculation of the Bethe logarithm in
few-electron atoms. The first one starts with the definition (\ref{eq:55}) and uses the basis-set
representation of the Hamiltonian as a sum of the spectrum of the eigenfunctions. The difficulty
is that the sum in the numerator is nearly diverging because the dominant contribution comes
from the high-energy continuum states of the spectrum. This problem is solved by using a basis
set whose spectrum of pseudostates spans a huge range of energies \cite{drake:99:cjp}.

An alternative approach was first introduced by C.~Schwartz \cite{schwartz:61} and further
developed by V.~Korobov \cite{korobov:99,korobov:04,korobov:19:bethelog}. Within this method, the
Bethe logarithm $\ln k_0$ is represented as an integral over the momentum of the virtual photon,
with subtracting the ultraviolet asymptotics and performing the limit,
\begin{align} \label{eq:61}
\bl = \frac1D \lim_{\Lambda \to \infty} \Biggl[
  \bigl< {\vec{\nabla}}^2\bigr>\,\Lambda  + D\, \ln 2\Lambda  + \int_0^\Lambda  dk\, k\, J(k)
  \Biggr]\,,
\end{align}
where $D = 2\pi Z \bigl< \delta^3(r_1)+\delta^3(r_2)\bigr>$, $\vec{\nabla} \equiv
\vec{\nabla}_1+\vec{\nabla}_2$, and
\begin{align}\label{eq:63}
  J(k) = \Big< \vec{\nabla}\, \frac1{E_0-H_0-k}\, \vec{\nabla} \Big>\,.
\end{align}
The asymptotic expansion of $J(k)$ for large $k$ reads
\begin{align}\label{eq:63a}
  J(k) \underset{k\to \infty}{=} -\frac1k\,\lbr \vec{\nabla}^2\rbr -\frac{D}{k^2} + \frac{2\sqrt{2} Z D}{k^{5/2}} - 2Z^2D\frac{\ln k}{k^3}+ \ldots\,.
\end{align}
Splitting the integration interval $(0,\Lambda)$ into two parts $(0,K)$ and $(K,\Lambda)$, where
$K$ is an arbitrary cutoff parameter, we can rewrite Eq.~(\ref{eq:61}) as
\begin{align} \label{eq:64}
\bl = & \ \frac{K}{D}\,\lbr \vec{\nabla}^2\rbr + \ln(2K)
 + \frac1D \int_0^K  dk\, k\, J(k)
 \nonumber \\ &
  + \frac1D \int_K^{\infty}  dk\, k\,\Big[ J(k) + \frac1k\,\lbr \vec{\nabla}^2\rbr +\frac{D}{k^2}\Big]
  \,.
\end{align}
The above expression is finite, does not depend on $K$, and is suitable for a numerical
evaluation.

We now address the angular reduction in the second-order matrix element $J(k)$ given by
Eq.~(\ref{eq:63}). It is performed in several steps. First, we represent the gradient acting on
the reference-state wave function $\nabla^j\,\psi^{i_1 .. i_L}$ as a sum of irreducible Cartesian
tensors, as described in Appendix~{\ref{app:irreducible}}. For example, the gradient acting on a
$P$-state wave function $\nabla^j\,\psi^{i}$ is represented as a sum of the $L = 0$, $L = 1$, and
$L = 2$ irreducible Cartesian tensors, which induce, correspondingly, the $L = 0$, $L = 1$, and
$L = 2$ angular-momentum contributions from the resolvent. The second-order matrix element of an
irreducible tensor $\Phi^{i_1 .. i_L}$ is transformed as
\begin{align}
&\ \Big< \Phi^{i_1 .. i_L} \Big| \frac1{E_0-H_0-k} \Big| \Phi^{i_1 .. i_L} \Big>
&= \lbr \Phi^{i_1 .. i_L} | \widetilde{\Phi}^{i_1 .. i_L}\rbr\,,
\end{align}
where $\widetilde{\Phi}^{i_1 .. i_J}$ is the solution of the inhomogeneous Schr\"odinger equation
\begin{align}
(E_0-H_0-k)\,\widetilde{\Phi}^{i_1 .. i_L} = \Phi^{i_1 .. i_L}\,.
\end{align}
Inserting the explicit representation of $\widetilde{\Phi}$ as a sum over the spectrum, we obtain
\begin{align} \label{eq:64bb}
&\ \Big< \Phi^{i_1 .. i_L} \Big| \frac1{E_0-H_0-k} \Big| \Phi^{i_1 .. i_L} \Big>
=
\sum_n \frac{\Big|\lbr \Phi^{i_1 .. i_L} | \psi_n^{i_1 .. i_L}\rbr\Big|^2}{E_0-E_n-k}\,.
\end{align}
An alternative way to arrive at this expression is to observe that the scalar product
$\langle\Phi|\psi\rangle$ includes an integration over the continuous and a summation over the
discreet variables, namely $\langle\Phi|\psi\rangle \equiv \langle \Phi^{i_1..i_L} |
\psi^{i_1..i_L}\rangle = \sum_{i_1 .. i_L}\int d^{3n} r \,
\Phi^{i_1..i_L*}(r)\,\psi^{i_1..i_L}(r)$.

The advantage of the integral representation of the Bethe logarithm  is that $J(k)$ has a form of
the symmetric second-order perturbation correction and thus obeys the variational principle. We
therefore can variationally optimize the basis-set representation of the resolvent
$1/(E_0-H_0-k)$ for different $k > E_0-E_{(1s)^2}$. For lower values of $k$, the basis can be
variationally optimized if one fixes pre-optimized parameters of the more deeply bound states
with $E_n < E_0$.

Our numerical procedure was performed in two steps. First, we optimized the basis for several
different scales of the photon momentum, $k = 10^i$, with typical values of $i = 1,..,4$. After
that, the computation of the function $J(k)$ was performed with a basis obtained by merging
together the optimized sets for the two closest $k_i$ points, thus essentially doubling the size
of the basis. In the second step, we perform the integration over $k$. The integral over $(0,K)$
(with the typical choice of $K = 10$) was calculated analytically, after the full diagonalization
of the Hamiltonian matrix. The remaining interval was split into two parts, $(K,K_2)$ and
$(K_2,\infty)$, with the typical choice of $K_2 = 10^4$. The integral over the former was
performed with help of Gauss-Legendre quadratures, after the change of variables $t = 1/k^2$. The
remaining part of the integral was calculated analytically, after fitting numerical values of
$J(k)$ to the known form of the asymptotic expansion,
\begin{align}
  J(k) = {\rm pol}\left(\frac1{\sqrt{k}}\right) +
  \frac{\ln k}{k}\,{\rm pol}\left(\frac1k\right)\,,
\end{align}
where ${\rm pol}(x)$ denotes a polynomial of $x$. The first terms of this expansion are given by
Eq.~(\ref{eq:63a}), whereas the higher-order coefficients are obtained by fitting.

Calculations of the Bethe logarithm for the finite nuclear mass can be performed analogously to
the above, or by perturbation theory. The numerical procedure for evaluation of the
recoil correction to the Bethe logarithm by perturbation theory is described in Appendix~A of
Ref.~\cite{yerokhin:10:helike}.

Table~\ref{tab:BL} presents a comparison of different calculations of the Bethe logarithm for the
$2^3P$ state of helium. The most accurate results for the ground and excited states of helium are
obtained by Korobov in Ref.~\cite{korobov:19:bethelog}. Results for He-like ions can be found in
Refs.~\cite{drake:99:cjp,yerokhin:10:helike}.

\begin{table}
\caption{Comparison of different calculations of the Bethe logarithm for the $2^3P$ state of helium.
\label{tab:BL}}
\begin{tabular}{l w{5.15}}
\hline
Drake and Goldman 1999 \cite{drake:99:cjp}            & 4.369\,985\,20\,(2)\\
Korobov 2004 \cite{korobov:04}                        & 4.369\,985\,356\,(1) \\
Yerokhin and Pachucki 2010 \cite{yerokhin:10:helike}  & 4.369\,985\,364\,4\,(2)  \\
Korobov 2019 \cite{korobov:19:bethelog}               & 4.369\,985\,364\,549\,(3) \\
\hline
\end{tabular}
\end{table}

\section{$\bm{\alpha^6\,m}$ QED effects}

The $\alpha^6\,m$ QED corrections to energy levels in atoms are represented by the sum of the
expectation value of the effective $\alpha^6\,m$ Hamiltonian $H^{(6)}$ and the second-order
perturbation correction induced by the Breit Hamiltonian,
\begin{align}\label{eq:66}
E^{(6)} = \lbr H^{(6)}\rbr + \Big< H_{{\rm Breit}, R}^{(4)} \, \frac{1}{(E_0-H_0)'} \, H_{{\rm Breit}, R}^{(4)} \Big>\,,
\end{align}
where $H_{{\rm Breit}, R}^{(4)}$ is the regularized Breit Hamiltonian of the order $\alpha^4\,m$,
\begin{align}
H_{{\rm Breit}, R}^{(4)} = H_{AR} + H^{(4)}_B + H^{(4)}_C + H^{(4)}_D\,.
\end{align}
We note that in order to avoid admixture of higher-order contributions in $E^{(6)}$, we have to
retain only the $\alpha^4\,m$ part in the definition of the Breit Hamiltonian, i.e., to set the
magnetic moment anomaly $\kappa\to 0$ in the definitions (\ref{eq:36})-(\ref{eq:38}). This is indicated by
the superscript ``4'' in the corresponding operators.

Formulas for the effective $\alpha^6\,m$ Hamiltonian $H^{(6)}$ are rather lengthy and will not be
reproduced here. In the case of fine structure, they were first obtained by Douglas and Kroll in
1974 \cite{douglas:74} and later re-derived in Refs.~\cite{zhang:96:a,pachucki:99:jpb}. For the
energy centroid, the situation is greatly complicated because of the appearance of numerous diverging
operators. The corresponding derivation was accomplished by one of us (K.P.), in
Ref.~\cite{pachucki:02:jpb} for the triplet states and in Ref.~\cite{pachucki:06:hesinglet} for
the singlet states of helium. The complete formulas suitable for numerical evaluation can be
found in Ref.~\cite{wienczek:19}.

The nuclear recoil $\alpha^6\,m$ correction has the same structure as the non-recoil one, but the
expressions for the operators are much more complicated. This correction was calculated in
Ref.~\cite{patkos:16:triplet} for the triplet states and in Ref.~\cite{patkos:17:singlet} for the
singlet states of helium.

\subsection{Second-order terms}

We now discuss the evaluation of the second-order contributions, represented by the second term
in Eq.~(\ref{eq:66}). Such corrections were first calculated for the fine structure by Hambro
\cite{hambro:72} and by Lewis and Serafino \cite{lewis:78}. Later, the fine-structure
calculations  were greatly improved in Refs.~\cite{yan:95:prl,drake:02:cjp,pachucki:02:jpb:a}.
For the centroid energies, the second-order corrections were calculated in
Refs.~\cite{pachucki:06:hesinglet,pachucki:06:he} for the $2S$ and $2P$ states and in
Refs.~\cite{wienczek:19,yerokhin:20:dstates} for the $nD$ states of helium.

It is convenient to rewrite Eq.~(\ref{eq:66}),
expressing the second-order perturbation correction more explicitly,
\begin{align}
E^{(6)} = &\ \lbr H^{(6)}\rbr + \Big< H_{AR} \, \frac{1}{(E_0-H_0)'} \, H_{AR} \Big>
 \nonumber \\ &\hspace*{-7ex}
  + \Big< H_{B}^{(4)} \, \frac{1}{(E_0-H_0)'} \, H_{B}^{(4)} \Big>
  + \Big< H_{C}^{(4)} \, \frac{1}{(E_0-H_0)'} \, H_{C}^{(4)} \Big>
 \nonumber \\ &\hspace*{-7ex}
  + \Big< H_{D}^{(4)} \, \frac{1}{(E_0-H_0)'} \, H_{D}^{(4)} \Big> + 2\,\Big< H_{B}^{(4)} \, \frac{1}{(E_0-H_0)'} \, H_{D}^{(4)} \Big>
 \nonumber \\ &
+ 2\,\Big< H_{AR} \, \frac{1}{(E_0-H_0)'} \, \big[ H_{B}^{(4)} + H_{D}^{(4)}\big] \Big>\,.
\end{align}
We note that the non-symmetrical second-order corrections (the last two terms in the above
equation) vanish for the centroid energy, but contribute to the fine structure.

The second-order perturbative corrections are calculated as follows. In the first step, we
perform traces over the spin degrees of freedom in the matrix elements. Then we
decompose the product of a tensor operator Q and the reference-state wave function $\psi$ into
the irreducible tensor parts $\tilde\psi$, as described in Appendix~\ref{app:irreducible}. In the
last step we calculate the second-order matrix elements induced by the irreducible parts
$\tilde\psi$ as (see Eq.~(\ref{eq:64bb}))
\begin{align}
\Bigl\langle\tilde\psi\Bigl |\frac{1}{(E_0-H_0)'}\Bigr|\tilde\psi\Bigr\rangle =
\sum_n^{E_n\ne E_0} \frac{|\langle\tilde\psi^{i_1..i_J}|\psi_n^{i_1..i_J}\rangle|^2}{E_0-E_n}\,.
\end{align}

The numerical evaluation of symmetrical second-order contributions was carried out with the
variational optimization of the nonlinear parameters of the basis set for the resolvent
$1/(E_0-H_0)$. Convergence of numerical results is often rather slow, especially for
contributions with $H_{AR}$. This is associated with the fact that the effective wave function
$|\delta \psi\rbr = 1/(E_0-H_0)'|H_{AR}\rbr$ has an integrable singularity at $r_a\to 0$. In
order to represent such wave functions with the exponential basis, very large (both positive and
negative) exponents are required. In order to effectively span large regions of parameters, we
used non-uniform distributions of the nonlinear parameters. E.g., for the nonlinear parameters
$\alpha_i$ we used the distributions of the kind \cite{pachucki:02:jpb}
\begin{equation}
\alpha_i = A_1+ (t_i^{-a}-1)\,A_2\,,
\end{equation}
with $a = 2$ and 3, where the variable $t_i$ has a uniform quasirandom distribution over the
interval $(0,1)$ and the variables $A_{1,2}$ are subjects of variational optimization. An example
of the convergence study of the second-order correction $\lbr H_{AR}\, \frac1{(E_0-H_0)'}\,
H_{AR}\rbr$ is given in Table~\ref{tab:sec}. Numerical evaluation of non-symmetrical second-order
contributions was carried out with basis sets, optimized for the corresponding symmetrical
corrections.

\begin{table}
\caption{Convergence study of the second-order perturbation correction $\lbr H_{AR}\, \frac1{(E_0-H_0)'}\, H_{AR}\rbr$
for the $2^3P$ state of helium.
Units are $\alpha^6m = \alpha^4$\,a.u. \label{tab:sec}}
\begin{tabular}{c w{5.15}w{4.9}}
\hline
 \multicolumn{1}{c}{$N$}
 & \multicolumn{1}{c}{$\Delta E$}
         & \multicolumn{1}{c}{Increment}
                        \\\hline\\[-7pt]
  200 &  -15.847\,010\,059  &  \\
  400 &  -15.848\,416\,968  &  -0.14 \times 10^{-2} \\
  800 &  -15.848\,498\,832  &  -0.82 \times 10^{-4} \\
 1200 &  -15.848\,507\,251  &  -0.82 \times 10^{-5} \\
 1600 &  -15.848\,508\,295  &  -0.10 \times 10^{-5} \\
 2000 &  -15.848\,508\,667  &  -0.37 \times 10^{-6} \\
 2400 &  -15.848\,508\,705  &  -0.39 \times 10^{-7} \\
 2800 &  -15.848\,508\,781  &  -0.75 \times 10^{-7} \\
 Ref.~\cite{pachucki:06:he}
      &  -15.848\,510(2)  \\
\hline
\end{tabular}
\end{table}

\section{$\bm{\alpha^7\,m}$ QED effects}

The $\alpha^7\,m$ QED correction to energy levels in atoms is given \cite{pachucki:06:prl:he} by
the sum of the relativistic correction to the Bethe logarithm $E_L$, the expectation value of the
effective $\alpha^7\,m$ Hamiltonian $H^{(7)}$, and the perturbation of the $\alpha^5\,m$ QED
operator by the Breit Hamiltonian,
\begin{align}\label{eq:70}
E^{(7)} = E_L + \lbr H^{(7)}\rbr + 2\, \Big< H_{{\rm Breit}, R}^{(4)} \, \frac{1}{(E_0-H_0)'} \, H^{(5)}_R \Big>\,.
\end{align}
The regularized effective $\alpha^5\,m$ Hamiltonian is \cite{patkos:21}
\begin{align}
 H^{(5)}_R = &\ -\frac{Z}{\pi}\Big( \frac{19}{45} + \frac23\,\ln\frac{\alpha^{-2}}{2}\Big)
  \Big(\frac{\vec{r}_1\cdot\vec{\nabla}_1}{r_1^3} + \frac{\vec{r}_2\cdot\vec{\nabla}_2}{r_2^3}\Big)
 \nonumber \\ &
  -\frac{7}{6\pi}\frac1{r^3} + H_B^{(5)} + H_C^{(5)} + H_D^{(5)}\,,
\end{align}
where $H_{B,C,D}^{(5)}$ are the $O(\alpha)$ parts of the corresponding spin-dependent parts of
the Breit Hamiltonian, $H_B = H_B^{(4)} + \alpha\,H_B^{(5)}$, etc. The operator $H^{(5)}_R$ is
non-Hermitian and is assumed to act on a ket trial function $\phi$ on the right.

The relativistic correction to the Bethe logarithm is rather complicated. We will not discuss its
calculation here, but direct the reader to original studies. This correction was first calculated
for the fine structure of the $2^3P$ state; the corresponding calculations for helium and
helium-like ions were performed in
Refs.~\cite{pachucki:06:prl:he,pachucki:09:hefs,pachucki:10:hefs}. In our recent investigation
\cite{yerokhin:18:betherel} we performed a calculation for the energy centroid of the $2^3S$ and
$2^3P$ states. For singlet states of helium, this correction has never been calculated so far.

The derivation of the effective $\alpha^7\,m$ Hamiltonian $H^{(7)}$ for helium is an extremely
difficult problem. It was first accomplished by one of us (K.P.) for the fine structure in
Refs.~\cite{pachucki:06:prl:he,pachucki:09:hefs}. Recently, we performed
\cite{patkos:20,patkos:21} the derivation of $H^{(7)}$ for triplet states of helium and
calculated \cite{patkos:21:helamb} the corresponding correction to the energies of the $2^3S$ and
$2^3P$ states. For singlet states, the effective $\alpha^7\,m$ Hamiltonian is unknown.

From the computational point of view, the main difficulty of the evaluation of the $\alpha^7\,m$
correction is the calculation of the Bethe-logarithm contribution $E_L$. The computational scheme
is similar to that for the plain Bethe logarithm and is described in
Ref.~\cite{yerokhin:18:betherel}. Conversely, the computation of the expectation value of
$H^{(7)}$ and the second-order corrections is very similar to the calculation of the
$\alpha^6\,m$ corrections.

\section{Other effects}

The finite nuclear size correction is given by (in relativistic units)
\begin{align}
E_{\rm fns} = \frac{2\,\pi}{3}\,Z\,\alpha\, &\ \big< \delta^{(3)}(r_1)+ \delta^{(3)}(r_2)\big>\,R^2\,
 \nonumber \\ & \times
 \big[1-(Z\,\alpha)^2\,\ln(m\,R\,Z\,\alpha)\big]\,,
\end{align}
where $R$ is the root-mean-square nuclear charge radius, and the expectation value of the Dirac
$\delta$ functions is assumed to include the finite-nuclear-mass correction induced by $\delta_M
H$.

The higher-order QED effects are approximated on the basis of known results for hydrogenic atoms.
 Specifically, the hydrogenic one-loop and two-loop corrections
for the $2s$ state of He$^+$ are given by \cite{yerokhin:18:hydr}
\begin{align} \label{eq:ma8a}
E^{(8+)}_{\rm rad1}({\rm hydr}) =&\ \frac{Z^7}{8\pi}\,  83.824\,,
 \\
E^{(8+)}_{\rm rad2}({\rm hydr}) =&\ \frac{Z^6}{8\pi^2}\, \Big( -\frac8{27}\,\ln^3 [(Z\alpha)^{-2}]
 + 0.639\,\ln^2 [(Z\alpha)^{-2}] \nonumber\\ &\
+ 41.387\,\ln [(Z\alpha)^{-2}]  -81.1 \pm 10 \Big)\,.
 \label{eq:ma8b}
\end{align}
An approximation for the higher-order $\alpha^8m$ QED correction to the ionization energies of
the helium atom is obtained from the corresponding hydrogenic $2s$ contribution by
\begin{align} \label{eq:ma8c}
E^{(8+)} = E^{(8+)}({\rm hydr})\, \frac{\big< \delta^3(r_1) + \delta^3(r_2)\big> - \nicefrac{Z^3}{\pi}}{\nicefrac{Z^3}{8\pi}}\,.
\end{align}

\section{Comparison of theory and experiment}

\begin{table*}
\caption{Comparison of theory and experiment for the intrashell $n = 2$ transitions in $^4$He, in MHz. $2^3P$ stands for the
centroid energy. \label{tab:2L}
}
\begin{ruledtabular}
\begin{tabular}{l w{10.5}w{10.5} c w{3.7}}
\centt{Transition}
  & \centt{Theory \cite{patkos:21,pachucki:17:heSummary,wienczek:19}} & \multicolumn{1}{c}{Experiment}& \multicolumn{1}{c}{Reference} & \multicolumn{1}{c}{Difference} \\ \hline\\[-7pt]
$2\,^1\!S_0$ -- $2\,^1\!P_1$   & 145\,622\,891.6\,(2.3)  & 145\,622\,892.89\,(18)    & Luo 2013 \cite{luo:13}           & -1.3\,(2.3)\\
$2\,^3\!S_1$ -- $2\,^1\!P_1\,$ & 338\,133\,595.8\,(0.4)  & 338\,133\,594.4\,(5) \, &Notermans 2014 \cite{notermans:14}         &\ \ 1.4\,(0.6) \\
$2\,^3\!S_1$ -- $2\,^1\!S_0$   & 192\,510\,704.2\,(1.9)  & 192\,510\,702.148\,72\,(20) \, &Rengelink 2018 \cite{rengelink:18}           &\ \ 2.1\,(1.9)\\
$2\,^3\!S_1$ -- $2\,^3\!P\ $   & 276\,736\,495.620\,(54) & 276\,736\,495.600\,0\,(14)^a \, &Zheng 2017 \cite{zheng:17}      &   0.020\,(54)\\
                 &                         & 276\,736\,495.649\,(2)\,^a \, &Pastor 2004 \cite{pastor:04}                  &   -0.029\,(54)\\
\end{tabular}
$^a$ using theoretical results for the $2^3P$ fine structure from Table~\ref{tab:fs}.
\end{ruledtabular}
\end{table*}

\begin{table*}
\caption{Comparison of theory and experiment for the ionization energy (IE)
and $n$-$n'$ transitions in $^4$He, in MHz.  \label{tab:differentLs}
}
\begin{ruledtabular}
\begin{tabular}{l cc c c}
\centt{Transition}
  & \centt{Theory \cite{patkos:21,pachucki:17:heSummary,wienczek:19}} & \multicolumn{1}{c}{Experiment}& \multicolumn{1}{c}{Reference} & \multicolumn{1}{c}{Difference} \\ \hline\\[-7pt]
$1\,^1\!S_0$\,(IE)       & 5\,945\,204\,172\,(36)            & 5\,945\,204\,212\,(6)     & Kandula 2011 \cite{kandula:11} & $-$40\,(36)\\
$2\,^1\!S_0$\,(IE)       &    960\,332\,038.0\,(1.9)         &    960\,332\,041.01\,(15) & Lichten 1991 \cite{lichten:91}& $-$3.0\,(1.9) \\
                         &                                   &    960\,332\,040.491\,(32)& Clausen 2021 \cite{clausen:21}& $-$2.5\,(1.9) \\
$1\,^1\!S_0$ -- $2\,^1\!S_0$   &  4\,984\,872\,134\,(36) & 4\,984\,872\,315\,(48)        & Bergeson 1998 \cite{bergeson:98}   & $-$181\,(60) \\
$1\,^1\!S_0$ -- $2\,^1\!P_1$   &  5\,130\,495\,026\,(36) & 5\,130\,495\,083\,(45)        & Eikema 1997 \cite{eikema:98}      & $-$57\,(58)\\
$2\,^1\!S_0$ -- $3\,^1\!D_2$  & 594\,414\,289.3\,(1.9)   & 594\,414\,291.803\,(13)       & Huang 2018 \cite{huang:18}          &$-$2.5\,(1.9)\\
$2\,^1\!P_1$ -- $3\,^1\!D_2$  & 448\,791\,397.8\,(0.4)   & 448\,791\,399.11\,(27)        & Luo 2013 \cite{luo:13:b}            &$-$1.3\,(0.5)\\
$2\,^3\!S_1$ -- $3\,^3\!D_1$ & 786\,823\,849.540\,(57) & 786\,823\,850.002\,(56)         & Dorrer 1997 \cite{dorrer:97}      & $-$0.462\,(80)\\
$2\,^3\!P_0$ -- $3\,^3\!D_1$ & 510\,059\,754.863\,(28) & 510\,059\,755.352\,(28)         & Luo 2016 \cite{luo:16}   & $-$0.489\,(40)\\
\end{tabular}
\end{ruledtabular}
\end{table*}

\begin{table*}
\caption{Comparison of theory and experiment for the $2^3P$ fine-structure intervals in $^4$He, in kHz \label{tab:fs}. }
\begin{ruledtabular}
\begin{tabular}{ll  w{9.6}w{8.5}w{8.5}}
& \multicolumn{1}{l}{Reference}& \multicolumn{1}{c}{$2^3P_0 - 2^3P_2$} & \multicolumn{1}{c}{$2^3P_1 - 2^3P_2$} & \multicolumn{1}{c}{$2^3P_0 - 2^3P_1$} \\ \hline\\[-5pt]
Theory & \textrm{Pachucki and Yerokhin 2010} \cite{pachucki:10:hefs} & 31\,908\,131.4\,(1.7) & 2\,291\,178.9\,(1.7)  & 29\,616\,952.3\,(1.7) \\[5pt]
Experiment
&\textrm{Kato 2018 \cite{kato:18}} &   & 2\,291\,176.590\,(25)  \\
&\textrm{Zheng 2017 \cite{zheng:17}} &  31\,908\,130.98\,(13)   & 2\,291\,177.56\,(19)  \\
&\textrm{Feng 2015} \cite{feng:15} & & 2\,291\,177.69\,(36) \\
&\textrm{Smiciklas 2010} \cite{smiciklas:10}  $^a$  &  31\,908\,131.25\,(32) & \\
&\textrm{Borbely 2009} \cite{borbely:09} $^a$ & & 2\,291\,177.55\,(35) \\
&\textrm{Zelevinsky 2005} \cite{zelevinsky:05} $^a$ &  31\,908\,126.8\,(3.0) & 2\,291\,176.8\,(1.1) & 29\,616\,951.7\,(3.0) \\
&\textrm{Guisfredi 2005} \cite{giusfredi:05} $^a$ & &     & 29\,616\,953.\,(10.0)\\
&\textrm{George 2001} \cite{george:01} $^a$ &  &                         & 29\,616\,950.8\,(0.9) \\
&\textrm{Castillega 2000} \cite{castillega:00} $^a$ & &  2\,291\,177.1\,(1.0) \\
\end{tabular}
\end{ruledtabular}
$^a$ reevaluated in Refs.~\cite{marsman:15,marsman:15b}.
\end{table*}

In this section we summarize numerical results of QED calculations of energy levels in $^4$He and
compare theoretical predictions with available experimental results. Table~\ref{tab:2L} presents
such a comparison for transitions between states with the principal quantum number $n = 2$. We
note that our present theoretical uncertainty for the $2\,^3\!S\,$--$\,2\,^1\!S$ transition is
increased as compared to our previous work \cite{pachucki:17:heSummary}. The reason is an
accidental cancelation of the estimated $\alpha^7m$ term between the $2\,^3\!S$ and $2\,^1\!S$
states in Ref.~\cite{pachucki:17:heSummary}. Now the $\alpha^7m$ correction is calculated for the
$2\,^3\!S$ state and the theoretical uncertainty is defined by the $2\,^1\!S$ state only.
Table~\ref{tab:2L} shows good agreement of theory and experiment for the singlet-singlet and
triplet-triplet transitions but some tension for the singlet-triplet transitions. Specifically,
we note a 2.3$\,\sigma$ deviation from the experimental result \cite{notermans:14} for the
$2\,^3\!S$--$2\,^1\!P$ transition (with $\sigma$ denoting the standard deviation).

Of particular importance is the agreement observed for the $2\,^3\!P\,$--$\,2\,^3\!S$ transition,
because in this case two triplet states are involved, for which the theoretical accuracy is the
highest. Theoretical calculations of energies for the $2\,^3\!S$ and $2\,^3\!P$ states
\cite{patkos:21} are complete through order $\alpha^7\,m$, with resulting theoretical uncertainty
below 100~kHz, whereas for the $2\,^1\!S$ and $2\,^1\!P$ states the theory
\cite{pachucki:17:heSummary} is complete up to order $\alpha^6\,m$ only and the theoretical
accuracy is on the level of 1~MHz. For the $D$ states, theoretical calculations
\cite{wienczek:19,yerokhin:20:dstates} are also complete at order $\alpha^6\,m$, but the absolute
theoretical precision is much higher since the QED effects are smaller. In general, we conclude
that for the intrashell $n=2$ transitions there is good agreement for transitions between the
states with the same spin multiplicity and some tension for the states of different spin
multiplicity.

The situation becomes even more strained when we consider ionization energies and transitions
involving states with different $n$'s. The corresponding comparison is presented in
Table~\ref{tab:differentLs}. We immediately notice that all differences between theory and
experiment are of the same sign and that most of them are outside of the theoretical error bars.
The largest discrepancies are found for the $2\,^3\!S_1\,$--$\,3\,^3\!D_1$ and the
$2\,^3\!P_0\,$--$\,3\,^3\!D_1$ transition, of 6 and 12$\,\sigma$, correspondingly. These
transitions involve the triplet states, for which theoretical uncertainties are the smallest, so
that 0.5~MHz differences from the experimental values lead to large relative deviations.

The comparison in Tables~\ref{tab:2L} and \ref{tab:differentLs} suggests that there might be a
contribution missing in theoretical calculations of energy levels, which weakly depends on $L$
but strongly depends on the principal quantum number $n$ (the latter is natural because the
$1/n^3$ scaling is typical for all QED effects). This conjecture was put forward in
Ref.~\cite{yerokhin:20:dstates} and since then strengthened by subsequent calculations and
measurements. Such a missing contribution most likely originates from the $\alpha^6\,m$ or
$\alpha^7\,m$ QED corrections because all other theoretical effects are cross-checked against
independent calculations \cite{morton:06:cjp}.

Table~\ref{tab:fs} presents the comparison of theoretical and experimental results for the
fine-structure intervals of the $2^3P$ state in $^4$He. Theoretical predictions for these
intervals are of greater accuracy than for other intervals of the $n=2$ manifold. This is both
due to the fact that the theory of these intervals \cite{pachucki:10:hefs,pachucki:11} is
complete at the order $\alpha^7\,m$ and due to the smallness of QED effects. We observe a
generally good agreement between theory and experiment for the fine-structure intervals. The only
tension is a 1.4$\,\sigma$ deviation for the $P_{1,2}$ interval measured in Ref.~\cite{kato:18}.
We note that all pre-2010 experimental results were to a greater or lesser degree influenced by
unaccounted quantum-interference effects and were reevaluated in
Refs.~\cite{marsman:15,marsman:15b}.

Summarizing, we have reviewed a large amount of work accomplished during the last decades in
calculations of QED effects in the atomic structure of the helium atom. The leading-order
$\alpha^5\,m$ QED effects are nowadays well established by independent calculations and tested by
comparison with numerous experiments. However, recent calculations of higher-order $\alpha^6\,m$
and $\alpha^7\,m$ QED effects  revealed some small but systematic deviations from high-precision
experimental transition energies. Having in mind the importance of the helium spectroscopy for
determination of nuclear properties and fundamental constants, we conclude that further
theoretical and experimental efforts are needed in order to find the reasons behind the observed
discrepancies.

\section*{Author Contributions}
All authors contributed equally. All authors have read and agreed to the published version of the
manuscript.

\section*{Funding}
 V.A.Y. acknowledges support from the Russian Science Foundation
(Grant No. 20-62-46006). Work of V.P. and K.P. was supported by the National Science Center
(Poland) Grant No. 2017/27/B/ST2/02459. V.P. also acknowledges support from the Czech Science
Foundation - GA\v{C}R (Grant No. P209/18-00918S).

\section*{Conflicts of Interest}
The authors declare no conflict of interest.

\appendix

\section{Expectation values of singular operators}
\label{app:singularintegrals}

In this section we discuss the evaluation of matrix elements of singular operators $1/r^3$,
$1/r^4$, and $1/r^5$. The $1/r^3$ operator is standardly defined as
\begin{align} \label{1/r^3}
\int d^3 r\, \frac{f(\vec{r})}{r^3_\epsilon}
\equiv& \lim_{\epsilon\rightarrow0} \bigg[ \int_\epsilon^\infty d r \, \frac{f(r)}{r} + f(0)\,(\gamma_E+\ln \epsilon)\bigg]\,,
\end{align}
where $f(r) \equiv \int d\Omega\, f(\vec r)$ assumed to be a smooth function that allows a Taylor
expansion at $r = 0$. Further singular operators are defined \cite{patkos:20,patkos:21} as
\begin{align} \label{1/r^4}
\int d^3 r\, \frac{f(\vec{r})}{r^4_\epsilon}
\equiv& \lim_{\epsilon\rightarrow0}\bigg[ \int_\epsilon^\infty d r \frac{f(r)}{r^2} -\frac{f(0)}{\epsilon}+ f'(0)\,(\gamma_E+\ln \epsilon)\bigg]\,,\\
\int d^3 r\, \frac{f(\vec{r})}{r^5_\epsilon}
\equiv& \lim_{\epsilon\rightarrow0}\bigg[ \int_\epsilon^\infty d r \frac{f(r)}{r^3} -\frac{f(0)}{2\,\epsilon^2}-\frac{f'(0)}{\epsilon}
\nonumber \\ &
+ \frac{f''(0)}{2}\,(\gamma_E+\ln \epsilon)\bigg]\,,
\end{align}
and
\begin{align}
\int d^3 r\, f(\vec{r})\,\frac{\ln r}{r^4_\epsilon}
\equiv& \lim_{\epsilon\rightarrow0} \bigg[ \int_\epsilon^\infty d r \, \frac{f(r)\,\ln r}{r^2} - f(0)\,\frac{(1+\ln \epsilon)}{\epsilon}
\nonumber \\ &\ - f'(0)\,\frac{\ln^2 \epsilon}{2}\bigg]\,.\label{eq:E3}
\end{align}
Here we corrected the sign misprint in Eq.~(E3) of Ref.~\cite{patkos:21}. Note that the
definition of the $1/r^5$ operator given by Eqs.~(154) and (155) of Ref.~\cite{patkos:20} is
valid for triplet states only.

We now obtain explicit formulas for integrals of singular operators with exponential functions,
starting with the operator $1/r^3$,
\begin{align}\label{eq:b3}
I_{ \epsilon}(1,1,-2) = \lim_{\epsilon\rightarrow 0}\,
&\frac1{16\pi^2} \int d^3r_1 \int d^3r_2 \,e^{-\alpha r_1 -\beta r_2 - \gamma r}\,
 \nonumber \\ & \times
\biggl[\frac{1}{r^3}\,\Theta(r-\epsilon) + 4\,\pi\,\delta^3(r)\, (\gamma_E+\ln \epsilon)\biggr]\,.
\end{align}
It is evident that
\begin{align}
-\frac{\partial}{\partial \gamma} I_{ \epsilon}(1,1,-2) = I(1,1,-1)\,,
\end{align}
where $I(1,1,-1)$ can be immediately obtained from Eq.~(\ref{eq2:12c}). Therefore, the formal
integration of $I(1,1,-1)$ over the parameter $\gamma$ gives us an expression for $I_{
\epsilon}(1,1,-2)$, which is correct up to a $\gamma$-independent constant. The simplest way to
fix this constant is to examine the limit of Eq.~(\ref{eq:b3}) for $\gamma \to \infty$. For very
large $\gamma$, only the region of small $r$ contributes, and therefore
\begin{align}
&\,\frac1{16\pi^2} \int d^3r_1 \int d^3r_2 \frac{e^{-\alpha r_1 -\beta r_2 - \gamma r}}{r^3}\,
\Theta(r-\epsilon)
 \nonumber \\ & =
  2\int_{\epsilon}^{\infty}dr\,r \int_0^{\infty}dr_1\,r_1
  \int_{|r_1-r|}^{r_1+r}   dr_2\,r_2\, \frac{e^{-\alpha r_1 -\beta r_2 -\gamma
      r}}{r^3} \nonumber \\ & \approx \frac{2}{(\alpha+\beta)^3}\,
  \int_{\epsilon}^{\infty} dr \frac{e^{-\gamma r}}{r} \,.
\end{align}
Hence,
\begin{align}
  I_{\epsilon}(1,1,-2) \underset{\gamma\to\infty}{=}
  -\frac{2}{(\alpha+\beta)^3}\,\ln \gamma \,.
\end{align}
We conclude that the $\gamma$-independent constant in the limit $\gamma \to \infty$ vanishes. It
is interesting that this simple prescription holds also for other singular integrals. Fixing the
$\gamma$-independent constant, we arrive at the final result
\begin{align}\label{eq:b7}
I_{\epsilon}(1,1,-2)
= & \frac1{(\alpha+\beta)^3}\bigg[
-\ln \big[(\alpha+\gamma)(\beta+\gamma)\big] - \frac{8 \alpha \beta}{(\alpha-\beta)^2}
 \nonumber \\ &
 + \frac{(\alpha+\beta)^3 + 8 \alpha\beta\gamma}{(\alpha-\beta)^3}\, \ln \frac{\alpha+\gamma}{\beta+\gamma}
\bigg]\,.
\end{align}
We note that this expression has a spurious singularity at $\alpha = \beta$. It can be easily
removed if we separate the first two Taylor expansion terms of the logarithm function by
introducing $\overline{\ln}_2(x)$.

We now turn to the integral with $1/r^4$. Analogously to $1/r^3$, we write
\begin{align}
-\frac{\partial}{\partial \gamma} I_{ \epsilon}(1,1,-3) = I_{ \epsilon}(1,1,-2)\,.
\end{align}
So, integrating Eq.~(\ref{eq:b7}) over $\gamma$ and setting the $\gamma$-independent constant in
the limit $\gamma \to \infty$ to zero (this time we omit the justification), we obtain
\begin{widetext}
\begin{align}
I_{\epsilon}(1,1,-3)
= &
\frac{1}{{{({\alpha^2}-{\beta^2})}^3}}\Big\{2  (\alpha-\beta)  \big[\alpha  \beta  (\alpha+\beta)-({\alpha^2}-4  \alpha  \beta+{\beta^2})  \gamma\big]
\nonumber \\ &
-\big[2  \alpha  \beta
({\alpha^2}+{\beta^2})+{{(\alpha+\beta)}^3}  \gamma+4  \alpha  \beta  {\gamma^2}\big]  \ln \Big(\frac{\alpha+\gamma}{\beta+\gamma}\Big)\Big\}
+\frac{\gamma}{(\alpha+\beta)^3}  \, \ln [(\alpha+\gamma) (\beta+\gamma)]\,.
\end{align}
Repeating the same procedure once more, we obtain also a result for the $1/r^5$ integral,
\begin{align}
I_{\epsilon}(1,1,-4)
= &
\frac{1}{6  {{(\alpha^2-\beta^2)}^3} }
  \big[\alpha  \beta  {{(\alpha+\beta)}^3}+12 \alpha  \beta  ({\alpha^2}+{\beta^2})  \gamma+3
{{(\alpha+\beta)}^3}  {\gamma^2}+8  \alpha  \beta  {\gamma^3}\big]  \ln \Big(\frac{\alpha+\gamma}{\beta+\gamma}\Big)
\nonumber \\ &
+\frac{1}{18  {{(\alpha-\beta)}^2}  {{(\alpha+\beta)}^3}}\big[
-\alpha  \beta  (17  {\alpha^2}-10  \alpha  \beta+17  {\beta^2})-24  \alpha  \beta  (\alpha+\beta)  \gamma
+3 (9  {\alpha^2}-26  \alpha  \beta+9  {\beta^2})  {\gamma^2} \big]
\nonumber \\ &
+\frac{1}{6{{(\alpha+\beta)}^3}}   (\alpha  \beta-3  {\gamma^2})  \ln [(\alpha+\gamma) (\beta+\gamma)]\,.
\end{align}
\end{widetext}

\section{Tensor decomposition of a product of irreducible tensors}
\label{app:irreducible}
In calculations of the Bethe logarithm and the second-order perturbation corrections, we
encounter a problem of decomposition of products of irreducible Cartesian tensors into the
irreducible parts. In this section we collect formulas required for such decompositions. The
product of two vectors is represented as a sum of a symmetric and traceless second-rank tensor, a
vector, and a scalar,
\begin{align}\label{eq:B1}
P^i\,Q^j =&\ (P^i\,Q^j)^{(2)} + \frac{1}{2}\,\epsilon^{ijk}\,(\vec P\times\vec Q)^k+\frac{\delta^{ij}}{3}\,\vec P\cdot \vec Q\,.
\end{align}
The product of a vector and a symmetric and traceless second-rank tensor is decomposed as
\begin{align}
P^i\,Q^{jk} =&\ (P^i\,Q^{jk})^{(3)} + \epsilon^{ijl}\,T^{kl} + \epsilon^{ikl}\,T^{jl} \nonumber \\ &\
+\delta^{ij}\,T^k + \delta^{ik}\,T^j -\frac{2}{3}\,\delta^{jk}\,T^i\,, \label{A2}
\end{align}
where
\begin{align}
T^k =&\ \frac{3}{10}\,P^i\,Q^{ik}\,,\\
T^{kl} =&\ \frac{1}{6}\,P^i\,\bigl( \epsilon^{ijl}\,Q^{jk} + \epsilon^{ijk}\,Q^{jl}\bigr)\,.
\end{align}
This identity can be verified by contracting Eq.~(\ref{A2}) with $\delta^{ij}$ and
$\epsilon^{ijk}$. It can be easily extended to the higher-rank tensors $Q$.

Finally, we present the decomposition of the product of two symmetric and traceless tensors
$P^{ij}$ and $Q^{kl}$, required for calculations of second-order corrections involving
$D$-states,
\begin{widetext}
\begin{align}
P^{ij}\,Q^{kl} =&\  (P^{ij}\,Q^{kl})^{(4)} + \epsilon^{ika}T^{jal} + \epsilon^{jka}T^{ial} + \epsilon^{ila}T^{jak} + \epsilon^{jla}T^{iak} +
\delta^{ik}T^{jl}+\delta^{il}T^{jk} + \delta^{jk}T^{il}+\delta^{jl}T^{ik}
\nonumber \\ &\ - \frac{4}{3}\,\delta^{ij}T^{kl}-\frac{4}{3}\,\delta^{kl} T^{ij} +
                  T^a\,\big(\epsilon^{ika}\delta^{jl} + \epsilon^{ila}\delta^{jk} +
                  \epsilon^{jka}\delta^{il} + \epsilon^{jla}\delta^{ik}\big) +
                  T\,\big(\delta^{ik}\delta^{jl} + \delta^{il}\,\delta^{jk}-\frac{2}{3}\,\delta^{ij}\delta^{kl}\big)\,,
\end{align}
\end{widetext}
where
\begin{align}
  T^{jbl} =&\ \frac{1}{4}\,(\epsilon^{ikb}\,P^{ij}\,Q^{kl})^{(3)}\,,\\
  T^{jl} =&\ \frac{3}{7}\,(P^{ij}\,Q^{il})^{(2)}\,,\\
  T^b =&\ \frac{1}{10}\,\epsilon^{jlb}\,P^{ij}\,Q^{il}\,,\\
  T =&\ \frac{1}{10}\,P^{ij}\,Q^{ij}\,.
  \end{align}


\end{document}